\documentclass[namedreferences]{solarphysics}
\usepackage[hyperref,optionalrh,solaromanenum]{spr-sola-addons} % For Solar Physics 
\usepackage{graphicx}                    % For eps figures, newer & more powerfull
\usepackage{courier}                    % Change the \texttt command to courier style
\usepackage{color}                       % For color text: \color command
\usepackage[usenames,dvipsnames,svgnames]{xcolor}
\usepackage{lscape}     % For landscape tables and figures
\usepackage{pdflscape}  % So a landscape table can be read comfortably in a computer screen
\usepackage{rotating}
\usepackage{sidecap}

\usepackage[normalem]{ulem}     %Provides the following commands:
\newcommand{\etal}{{\it et al.}}

\renewcommand{\vec}[1]{ {\mathbf #1} }
\renewcommand{\vec}[1]{\mathbf{#1}}

\newcommand{\curl}{ {\bf \nabla} \times}

\newcommand{\BE}{\begin{equation}}
\newcommand{\EE}{\end{equation}}
\newcommand{\BA}{\begin{eqnarray}}
\newcommand{\EA}{\end{eqnarray}}
 \newcommand{\fig}[1]{Figure~\ref{fig:#1}}
 \newcommand{\figs}[2]{Figures~\ref{fig:#1} and \ref{fig:#2}}
 \newcommand{\figss}[2]{Figures~\ref{fig:#1}--\ref{fig:#2}}
 \newcommand{\sect}[1]{Section~\ref{sec:#1}}
 \newcommand{\sects}[2]{Sections~\ref{sec:#1} and \ref{sec:#2}}
 \newcommand{\sectss}[2]{Sections~\ref{sec:#1}--\ref{sec:#2}}

\newcommand{\arcsec}{''} 
\newcommand{\degree}{^{\circ}} 
\newcommand{\eg}{\textit{e.g.}}
\newcommand{\ie}{\textit{i.e.}}

\newcommand{\etc}{\textit{etc.}}

\begin{document}

\begin{article}

\begin{opening}

\title{\bf{Physical processes involved in the EUV ``Surge" Event of 09 May 2012}}

\author[addressref={1},corref,email={lopezf@iafe.uba.ar}]{\inits{M.}\fnm{Marcelo }\lnm{L\'opez Fuentes}
\orcid{0000-0001-8830-4022}}
\author[addressref={1,2},corref,email={}]{\inits{C. H. }\fnm{Cristina H. }\lnm{Mandrini}
\orcid{0000-0001-9311-678X}}
\author[addressref={1},corref,email={}]{\inits{M.}\fnm{Mariano }\lnm{Poisson}\orcid{0000-0002-4300-0954}}
\author[addressref={3},corref,email={}]{\inits{P.}\fnm{Pascal }\lnm{D\'emoulin}\orcid{0000-0001-8215-6532}}
\author[addressref={1,2},corref,email={}]{\inits{M.}\fnm{Germ\'an }\lnm{Cristiani}\orcid{0000-0003-1948-1548}}
\author[addressref={4},corref,email={}]{\inits{F. M.}\fnm{Fernando M. }\lnm{L\'opez}}%\orcid{}}
\author[addressref={1},corref,email={}]{\inits{M.}\fnm{Maria Luisa }\lnm{Luoni}\orcid{0000-0002-2825-7719}}

\runningauthor{M. L\'opez Fuentes \etal }
\runningtitle{Analysis of the ``Surge" of 09--May--2012}

\address[id={1}]{Instituto de Astronom\'\i a y F\'\i sica del Espacio (IAFE), CONICET-UBA, Buenos Aires, Argentina}

\address[id={2}]{Facultad de Ciencias Exactas y Naturales (FCEN), UBA, Buenos Aires, Argentina}

\address[id={3}]{LESIA, Observatoire de Paris, Universit\'e PSL, CNRS, Sorbonne Universit\'e, Univ. Paris Diderot, Sorbonne Paris Cit\'e, 5 place Jules Janssen, 92195 Meudon, France  }

\address[id={4}]{Instituto de Ciencias Astron\'omicas, de la Tierra y del Espacio (ICATE), CONICET, San Juan, Argentina}

\begin{abstract}
We study an EUV confined ejection observed on 09 May 2012 in active region (AR) NOAA 11476. For the analysis we use observations in multiple wavelengths (EUV, X-rays, H$\alpha$, and magnetograms) from a variety of ground-based and space instruments. The magnetic configuration showed the presence of two rotating bipoles within the following polarity of the AR. This evolution was present along some tens of hours before the studied event and continued even later. During this period the magnetic flux of both bipoles was continuously decreasing. A minifilament with a length of $\approx 30 \arcsec$ lay along the photospheric inversion line of the largest bipole. The minifilament was observed to erupt accompanied by an M4.7 flare (SOL20120509T12:23:00). Consequently, dense material, as well as twist, was injected along closed loops in the form of a very broad ejection whose morphology resembles that of typical H$\alpha$ surges. We conclude that the flare and eruption can be explained as due to two reconnection processes, one occurring below the erupting minifilament and another one above it. This second process injects the minifilament plasma within the reconnected closed loops linking the main AR polarities. Analyzing the magnetic topology using a force-free model of the coronal field, we identify the location of quasi-separatix layers (QSLs), where reconnection is prone to occur, and present a detailed interpretation of the chromospheric and coronal eruption observations. In particular, this event, contrary to what has been proposed in several models explaining surges and/or jets, is not originated by magnetic flux emergence but by magnetic flux cancellation accompanied by the rotation of the bipoles. In fact, the conjunction of these two processes, flux cancellation and bipole rotations, is at the origin of a series of events, homologous to the one we analyze in this article, that occurred in AR 11476 from 08 to 10 May 2012.
\end{abstract}


\end{opening}

%-----------INTRODUCTION------------------------------------------------

\section{Introduction}
\label{sec:intro}

%\quad{\S\bf~Activity, ejecta, CMEs, jets, surges and sprays }\\
Solar activity produces a variety of ejecta such as coronal mass ejections (CMEs), jets, surges, sprays, \etc~ They are usually classified according to their evolution, observed size, geometry, associated energies, and masses involved, but also according to the instruments, and hence wavelengths, with which they have been observed \citep[\eg ,][]{Rust80}. 

%\quad{\S\bf~Surges and sprays properties}\\
Surges and sprays, in particular, have been originally identified in H$\alpha$ images as the eruption of cold chromospheric material into the corona. While the material ejected in sprays does not apparently come back to the coronal base, in surges the material decelerates and then falls down along the same magnetic structure through which it was originally ascending \citep{Roy73} and/or reaches the opposite end of the closed structure. Surges have typical velocities in the range of 50 to 200 km s$^{-1}$, they reach heights up to 2$\times$10$^5$ km, and have durations of a few tens of minutes \citep{Foukal04}. They are often, though not always, observed in coincidence with the occurrence of flares \citep{Schmieder88}. 

%\quad{\S\bf~ EUV surges and jets}\\
With the advent of extreme ultraviolet (EUV) instruments it became clear that these ejections were also observable in this range \citep{Schmahl81}. Multi-wavelength analysis also demonstrated their connection with X-ray brightenings and jets \citep{Harrison90,Schmieder95,Canfield96}. A main difference between surges and jets is their magnetic configuration, which is closed or open, respectively. Jet like ejections observed in EUV have been named EUV jets by some authors \citep{Moore10,Liu04}. Observed in this band, the typical temperature of the ejected material is of a few 10$^5$ K; while densities have been found to be in the range from 10$^8$ to 10$^9$ cm$^{-3}$ \citep{Raouafi16}.

%\quad{\S\bf~Surges, jets and the photospheric field}\\
Concerning the photospheric magnetic field configuration where surges and jets occur, observations in the 1990s suggested that they could be indirectly originated due to the emergence of parasitic polarities, accompanied sometimes by flux cancellation \citep[see, \eg,][]{Schmieder95,Canfield96,Chae99}. High-resolution observations \citep{Jibben04,Brooks07,Guglielmino10,Uddin12,VargasDominguez14} also provide additional evidence for the close and repeated relation between flux emergence, chromospheric cold plasma ejections, and hot jets. Other investigations have shown that a combination of flux emergence and cancellation can play a role in the triggering of different types of jets \citep{Liu11,Panesar16}, while in others no flux emergence is observed \citep{Young14b,Young14a,Chandra17}.

%{\S\bf~First emergence models and other approaches}\\
Based on several of the observations discussed in the previous paragraph and the pioneer flux emergence model by \citet{Heyvaerts77}, \citet{Shibata92} and \citet{Yokoyama95,Yokoyama96} proposed a 2.5D numerical model in which the emergence of magnetic flux in an initial uniform coronal field could lead, via reconnection, to the ejection of cold plasma close to a hot jet. These authors described surges as resulting from a ``sling-shot effect due to reconnection, which produces a whip-like [plasma] motion''. \citet{Canfield96}, following the analysis of a series of H$\alpha$ surges and related X-ray jets, proposed a phenomenological 2D model based on reconnection to explain the observed evolution.

%{\S\bf~Recent models with emergence and with emergence and radiative transfer}\\
More recently, \citet{MorenoInsertis13} found in their 3D MHD simulation of an emerging region, the presence of a domain of dense cool plasma. \citet{MacTaggart15} performed 3D MHD simulations of the emergence of small-scale ARs in the presence of different ambient field configurations to determine where reconnection occurred and the characteristics of the flow of the dense plasma in different observed events. To understand the physical processes occurring in the photosphere, chromosphere, and corona during surges, \citet{Nobrega16} performed a 2.5D radiative-MHD numerical simulation of the emergence of a twisted magnetic tube in a statistically stationary magneto-convection configuration formed by the uppermost layers of the convective zone up to the corona. During the evolution of the system part of the emerged cold and dense material is ejected as in a surge.

%{\S\bf~Other simulations about the driver}\\
In the previously discussed models and simulations magnetic reconnection is driven by flux emergence; however, there are other simulations mainly applied to jets in which reconnection is forced by imposing horizontal photospheric twisting motions to a magnetic field configuration that includes a coronal magnetic null-point \citep{Pariat09,Pariat10,Pariat15,Pariat16}.

%{\S\bf~Jets observations including minifilaments}\\
Recent observations of coronal jets, either in coronal holes or in ARs, have identified the presence and eruption of small-scale filaments, called minifilaments, being part of the ejected material \citep{Shen12,Adams14,Sterling15,Sterling16,Panesar16b,Joshi18,Yang18,Moore18}. In another example, where lower resolution observations were analyzed, the presence of a constantly reformed minifilament and its eruption was postulated as the origin of a series of blow-out jets and the chain of consequent events \citep[flares and narrow CMEs,][] {Chandra17}.  
In the observed examples, the mechanism associated to the destabilization of the minifilament was the cancellation of magnetic flux along the polarity inversion line where it lay. Magnetic reconnection below the minifilament was responsible for an observed flare, while the same mechanism above the minifilament favoured the injection of its material into open field lines to form the jet. 

%{\S\bf~Simulations with minifilaments}\\
The identification of minifilament eruptions as the main origin of the plasma ejected in jets led \citet{Wyper17} and \citet{Wyper18} to propose that these mass ejections are produced by a break-out mechanism similar to what has been proposed to explain larger events like CMEs \citep[see, \eg,][]{Karpen12}. In these simulations, an emerged bipole is embedded in a unipolar uniform ambient field where a 3D coronal null-point is present. A small flux rope or minifilament is formed by shearing motions that also provide energy to the system. This strongly sheared field expands towards the null point and a reconnection process occurs which, like in the standard breakout CME model, removes the field restraining the small flux rope. Simultaneously, reconnection occurs below the small twisted flux rope as proposed in the articles mentioned in the previous paragraph, further building the flux rope. Finally, reconnection between the twisted flux rope and the background open field launches a jet.

%{\S\bf~Null}\\
In several magnetic configurations associated to jets and surges, H$\alpha$ brightenings at the sites of the events have a circular shape that can be closed \citep[or almost closed, see, \eg,][]{Joshi15,Sterling16,Li18}. In magnetic field models where a null point is found, the intersection of the separatrix (fan of the null) with the photospheric or chromospheric plane has a typical circular-like shape that agrees with observed chromospheric brightenings \citep[see, \eg,][for several examples]{Mandrini15,Masson12,Masson17}. Null-point topology is also supported by a large number of jet observations \citep[see the review by ][ and references therein]{Raouafi16}. 

%{\S\bf~QSLs}\\
Other magnetic topologies calculated from coronal extrapolation models have been found associated with jet and surge observations. Quasi-separatix layers \citep[QSLs, see, \eg,][for the original definition]{Demoulin96} were identified by \citet{Mandrini96} and \citet{Guo13}, with no magnetic null point present, in the case of jets. H$\alpha$ surges were observed in magnetic configurations having QSLs by \citet{Cristiani07} and in other configurations having bald patches by \citet{Mandrini02}. The latter topological structures where also observed during a series of blow-out jets and associated narrow CMEs  by \citet{Chandra17}. 
Finally, the examples with and without a magnetic null point are not conceptually different when considered within the QSL point of view, because a separatrix is only the extreme case when the field line mapping is becoming discontinuous in the QSL core.

%{\S\bf~Aim of this paper }\\
The event studied in this article is much less collimated than typical blow-out and EUV jets. Furthermore, like in surges, the material is ejected upwards, stays enclosed in the magnetic structures of the active region, and falls back along those structures. We will henceforth refer to it as an EUV ``surge" event, as it is most prominent in this wavelength range. We use quotation marks to indicate that it also bears differences with the classical surges analyzed elsewhere (see references at the beginning of this section). This EUV event occurred in active region (AR) NOAA 11576 on 09 May 2012 and is associated to an M4.7 flare with a peak in GOES soft X-ray light curve at $\approx$ 12:32 UT. Preliminary results of our analysis were discussed elsewhere \citep{LopezFuentes15}.

%{\S\bf~Roadmap}\\
In \sectss{obs_B}{sxr_obs} we study the evolution of the event in several wavelengths using data from the instruments described in \sect{data}. After modeling the AR magnetic field and computing its coronal topology, in \sect{topology} we interpret the role of QSLs in the magnetic configuration and we propose a phenomenological explanation of the event in terms of the %presence and 
eruption of the observed minifilament and its interaction with the surrounding magnetic structure. In \sect{conclusions} we discuss our results and present concluding remarks.

%-----------DATA DESCRIPTION AND OBERVATIONS-------------------

\section{Flare and EUV ``Surge'' Observations}
\label{sec:obs}

%---------------------------------------
\subsection{Data Description}
\label{sec:data}

%\quad{\S\bf~Instruments used and references}\\
To analyze the events on 09 May 2012, we use EUV data from the {\it Atmospheric Imaging Assembly} \citep[AIA:][]{Lemen12}, onboard the {\it Solar Dynamics Observatory} (SDO), and from the {\it Sun-Earth Connection Coronal and Heliospheric Investigation}  \citep[SECCHI:][]{Howard08}, onboard the {\it Solar Terrestrial Relations Observatory} (STEREO) spacecraft B, H$\alpha$ data from the {\it H-alpha Solar Telescope for Argentina} \citep[HASTA:][]{Bagala99,FernandezBorda02}, soft X-ray data from the {\it X-ray Telescope} \citep[XRT:][]{Golub07} onboard {\it Hinode}, hard X-ray data from the {\it Reuven Ramaty High Energy Solar Spectroscopic Imager} \citep[RHESSI:][]{Lin02}, and magnetograms from the {\it Helioseismic and Magnetic Imager} \citep[HMI:][]{Scherrer12}, onboard SDO.

%{\S\bf~Description of the data}\\
EUV data from SDO/AIA correspond to the 304 \AA~and 171 \AA~channels (henceforth, AIA 304 and 171). We select, from full disk images, subimages containing AR 11576 for the temporal range corresponding to the analyzed event. Coaligning these images to compensate for solar rotation we construct the EUV movies that accompany this article (\sect{obs_euv}). The images used for the movies are represented in logarithmic intensity scale for better contrast. 
We complement the SDO/AIA data with observations from the 195 \AA~channel of the SECCHI instrument on board STEREO-B. We use the 195 \AA~band because it has the highest temporal resolution (5 minutes) among the SECCHI channels. On 09 May 2012, STEREO-B spacecraft was located at an Earth ecliptic (HEE) longitude of -118$\degree$ away from Earth, from this location AR 11576 is seen at the solar limb. 

\begin{figure}[]
\begin{center}
\includegraphics[width=0.9\textwidth]{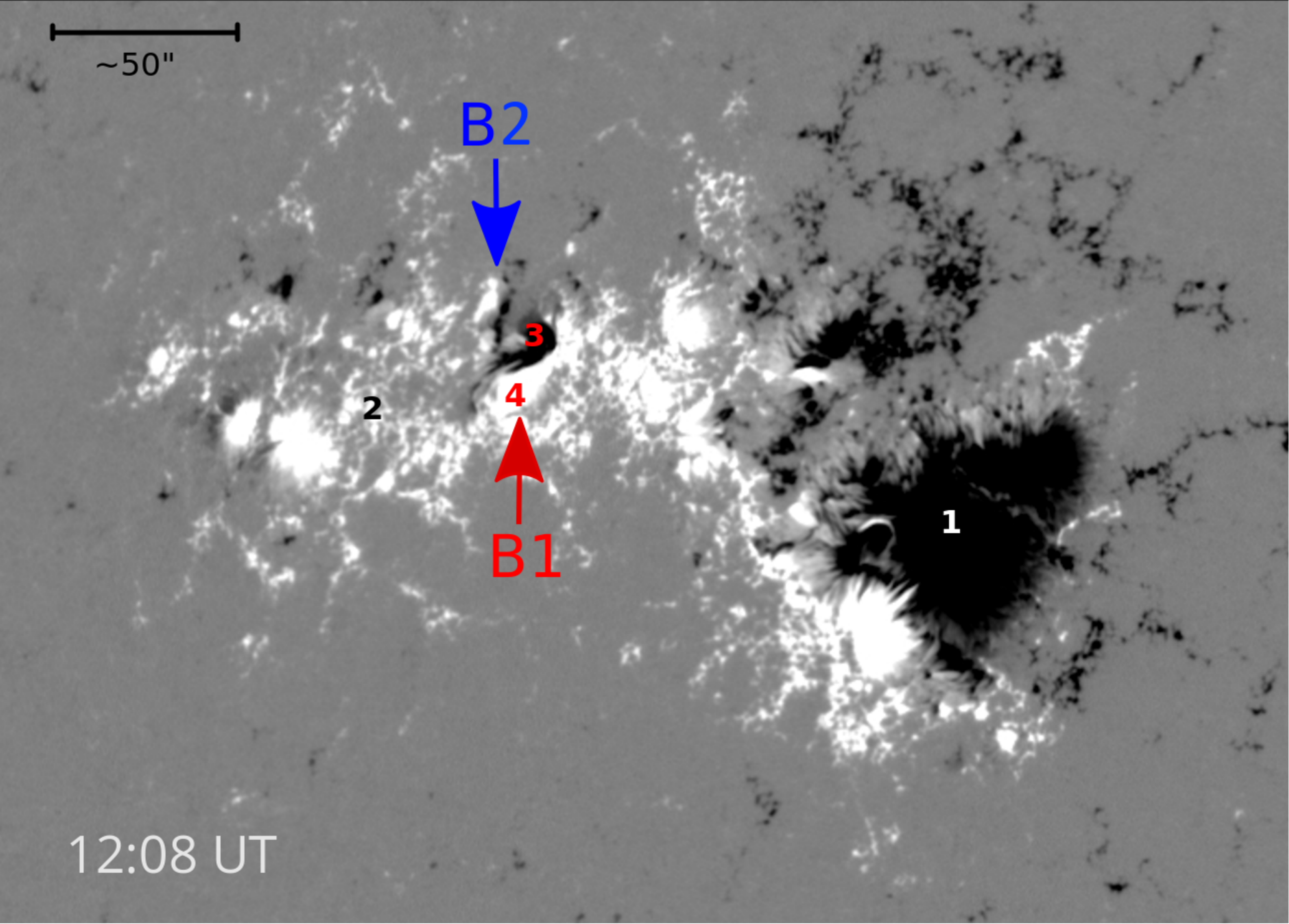}
\caption{SDO/HMI line-of-sight magnetogram of AR 11476, located at N12 E26, on 09 May 2012 at 12:08 UT. The arrows indicate the locations of the two rotating bipoles (B1 and B2) described in \sect{obs_B}. The flare and related ``surge" originated in their vicinity. The polarities involved in the studied event are numbered: 1 and 2 for the AR polarities, 3 and 4 for those of bipole B1. Black corresponds to negative magnetic field (pointing away from the observer) and white corresponds to positive field (towards the observer), the values of the field have been saturated above (below) 500 G (-500 G). The horizontal and vertical sizes of this image are 350\arcsec and 250\arcsec , respectively. On the top left, a black segment has been added to the panel to indicate the approximate scale size. A movie showing the magnetic field evolution along three days (\href{run:./HMI.mp4}{HMI.mp4}) is attached as electronic supplementary material.}
\label{fig:hmi}
\end{center}
\end{figure}

The analyzed SDO/HMI data consist of line-of-sight (LOS) magnetograms of AR 11576, selected from full-disk data. These magnetograms are used to study the evolution of the AR magnetic field, as described in \sect{obs_B}, and as boundary condition for the model described in \sect{model}.

The H$\alpha$ data from HASTA have a resolution of approximately 2\arcsec and are automatically taken in two different modes: normal (or patrol) and flare. In normal mode, during routine observations, the telescope takes full images once every 1 to 5 minutes. When a flare begins the instrument changes to flare mode and images are taken up to a rate of 2 {\it per} s. The data set used here consists of images taken in both modes depending on the time of observation, normal for before and after the analyzed flare and flare mode during the event (see the description of the timing in \sect{obs_chromos}). Similarly, {\it Hinode}/XRT images taken during the flare alternate between different filters (Ti-poly, Al-mesh, and Al-thick) and short or long exposure times. We combine these data in composite images in which the saturated pixels in the long-exposure images are replaced by the same corresponding pixels from the non-saturated short-exposure images. The XRT movie that accompanies this article has been made using these composite images (see \sect{sxr_obs}). Finally, RHESSI processed data correspond to hard X-ray counts in the channels of 
10\,--\,30 keV and above 40 keV, integrated during periods of 20 s at the time of the soft X-ray GOES flare emission peak (at $\approx$ 12:32 UT, see \sect{obs_chromos}). \\   

\begin{figure}[]
\begin{center}
\includegraphics[width=1.0\textwidth]{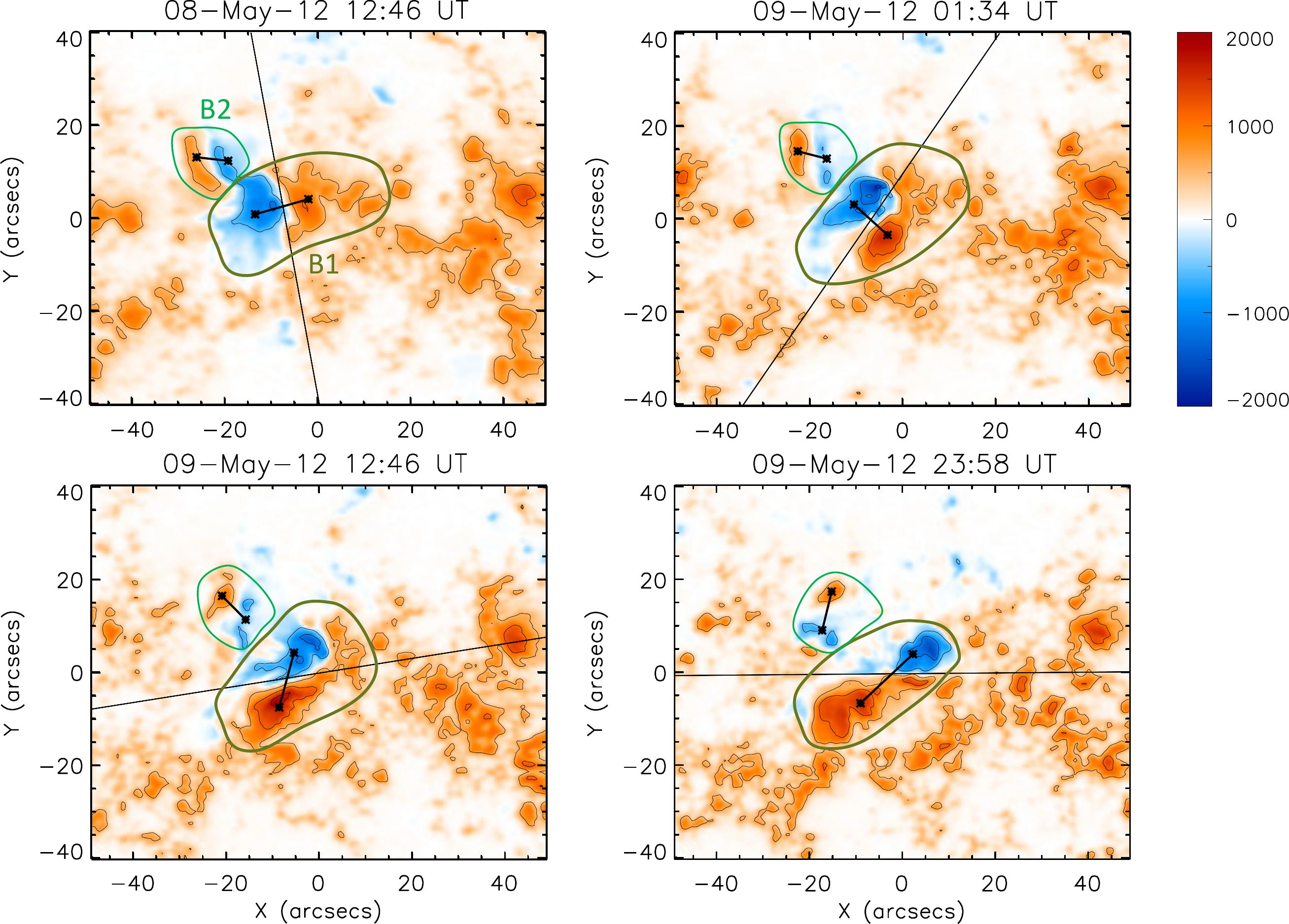}
\caption{The panels show four snapshots of the rotation of the bipoles B1 and B2. The dates and times are indicated at the top of each panel. The short black segments joining the asterisks (the positive and negative polarity barycenters) indicate the tilt angle. The long black straight line corresponds to the estimated mean polarity inversion line (PIL) of the largest bipole (B1). This bipole is enclosed by a green contour that surrounds the magnetic flux used in the computation of its polarity centers and rotation, as also done for B2. In these magnetograms the red (blue) areas correspond to positive (negative) magnetic field. Magnetic field contours of $\pm$ 1000, 1500, 2000 G are also included as reference. A movie included as electronic supplementary material shows the evolution of B1 (\href{run:./B1.mp4}{B1.mp4}).
}
\label{fig:tilt}
\end{center}
\end{figure}

%---------------------------------------
\subsection{Photospheric Magnetic Field Evolution}
\label{sec:obs_B} 

%\quad{\S\bf~Magnetic field configuration and pre-flare conditions}\\
AR 11476 appeared on the eastern solar limb on 04 May 2012. By 06 May it is clearly seen as having a global bipolar structure with a compact preceding negative polarity {\bf (numbered 1)} and a disperse following positive polarity {\bf (numbered 2)} that presents substantial magnetic flux fragmentation. A small bipole was located at the center of the positive polarity, well visible on 07 May. However, as the AR evolved, what looked as a single bipole appeared as two independent ones (see the area within the black box in the movie \href{run:./HMI.mp4}{HMI.mp4} included as supplementary material). The two bipoles have been identified as B1 and B2 (\fig{hmi}). The negative and positive polarities of B1 are identified as 3 and 4, respectively. The flare and mass ejection occurred in their vicinity.

An analysis of the evolution of B1 and B2 starting 36 h before the studied events indicates a rapid clockwise rotation of both bipoles, which implies a fast injection of magnetic helicity whose characteristics depend on the magnetic connectivities. This helicity injection is associated with  free magnetic energy accumulation and might also contribute to the minifilament  destabilization as proposed by \citet{Wyper17} and \citet{Wyper18}. The panels of \fig{tilt} show four different stages of the rotation of the bipoles. The full evolution can be seen in the movie that accompanies this article (\href{run:./B1.mp4}{B1.mp4}). 

%{\S\bf~PIL}\\ 
The rotation of each bipole implies also the rotation of their polarity inversion lines (PILs). The long black straight line in \fig{tilt} and in the corresponding movie is the PIL of B1, which we compute following the procedure developed by \citet{Poisson15}. The position of this straight line is defined to best separate the negative and positive polarities of the bipole B1. Following \citet{Poisson15,Poisson16}, the shape of the polarities at both sides of the PIL is an indication of a bipole formed by a negatively twisted flux rope.

%{\S\bf~Quantitative rotation}\\ 
To characterize quantitatively the rotation, we follow the evolution of the tilt angle of both bipoles. We first compute the positions of the barycenters of their positive and negative polarities (indicated by asterisks in \fig{tilt}) limiting the computation to the bipoles themselves, \ie~to the flux encircled by the green curve for B1 and B2, as shown in \fig{tilt}. Then, we define the tilt angle as the angle that the segment joining both barycenters forms with the solar equator direction. The evolution of the segment indicates that B1 rotates approximately 140$\degree$ clockwise during 36 h before the flare and continues rotating $\approx$ 40$\degree$ more until 10 May around 22:00 UT. In the case of B2, the rotation before the ejection and flare on 09 May is of $\approx$ 50$\degree$ clockwise, continuing for another 50$\degree$ in the same sense until it fully disappeared by $\approx$ 10:00 UT on 10 May.
 
%{\S\bf~B flux evolution}\\
 During the time period we measure the rotation of both bipoles, their magnetic flux decreased steeply as it cancelled with the surrounding magnetic field and probably also dispersed in the case of B2. In the case of B1, its negative flux is clearly seen decreasing against the positive flux concentration to the west (see \href{run:./HMI.mp4}{HMI.mp4}). As discussed in \sect{Processes}, this flux cancellation, together with B1 rotation, could play a key role in the triggering of the minifilament eruption. The magnetic flux of B1 decreased by around 90\% from 8 May until 10 May at $\approx$ 22:00 UT. 

\begin{figure}[]
\begin{center}
\includegraphics[width=0.9\textwidth]{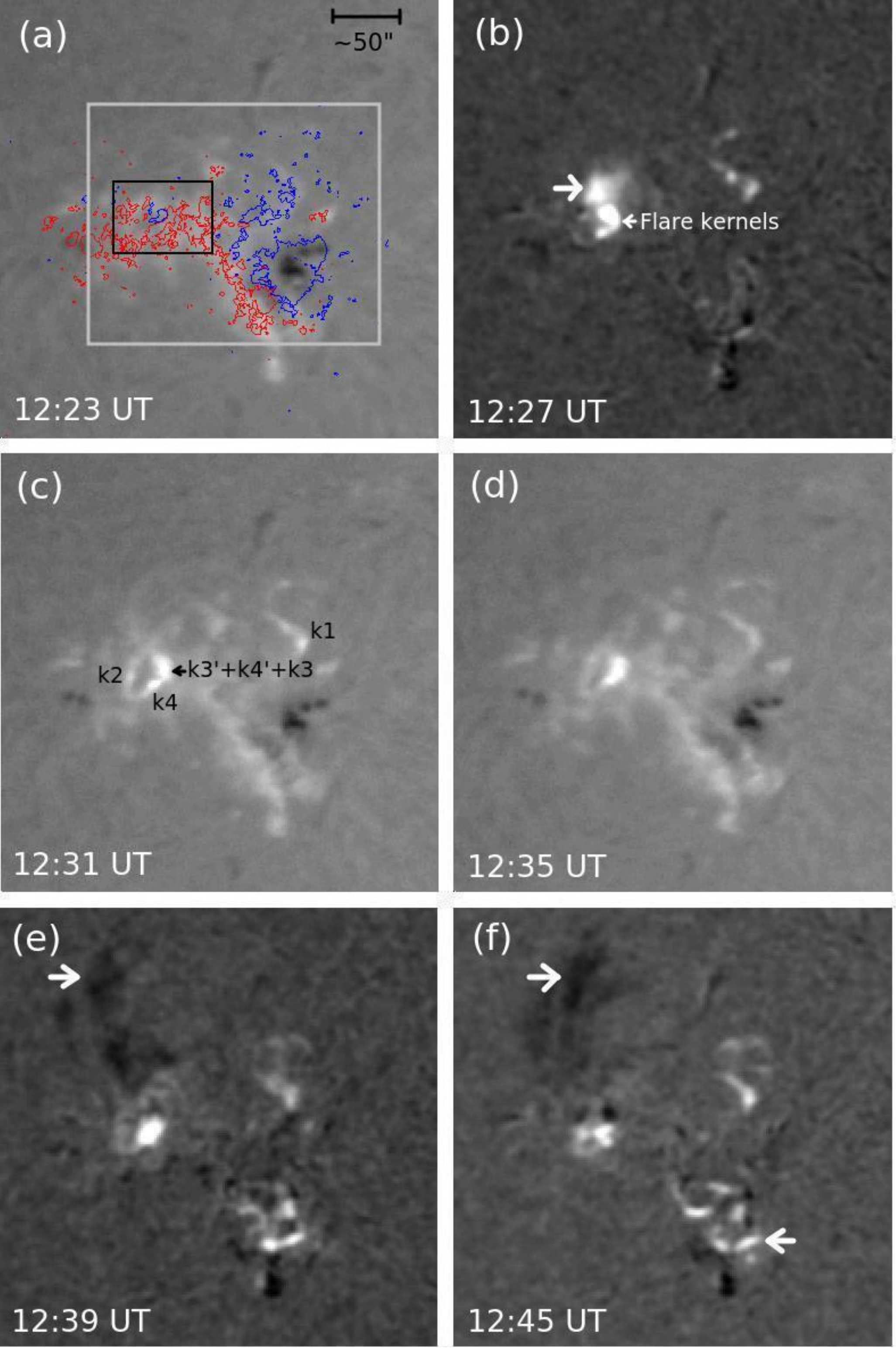}
\caption{HASTA images showing the flare evolution. The times are indicated at the bottom left of each panel. In panel a, $\pm$300 G HMI contours are included as reference. As in \fig{tilt} red (blue) corresponds to positive (negative) magnetic field. A black box is added to indicate the location of bipoles B1 and B2, it approximately covers the field of view of the panels in \fig{tilt}. 
The larger gray box indicates the FOV of \fig{closeup}. A black segment has been added at the top right to indicate the figure scale size. Panels b, e, and f correspond to base difference images (see \sect{obs_chromos}). The white arrow in panel b points to a north-eastern brightening that probably corresponds to heated ``surge'' plasmas it starts flowing upwards (see also the movie \href{run:./surge-closeup.mp4}{surge-closeup.mp4} that accompanies this article). In panel c, the flare kernels are labeled k1---k4 following the numbering used to label the magnetic polarities in \fig{hmi}. k3' and k4' indicate the kernels associated to the footpoints of loops that result from the internal reconnection process described in \sect{Processes}. The brightening pointed with a black arrow, and identified with the label k3'+k4'+k3, is composed by those three kernels (see also \fig{closeup}d). The upper white arrows in panels e and f show the location of the ``surge". The lower white arrow in panel f points to a second flare unrelated to the event studied in this article (see text in \sect{obs_chromos}).  A movie showing the H$\alpha$ evolution (\href{run:./HASTA.mp4}{HASTA.mp4}) is attached as electronic supplementary material. The square panels have a side length of approximately 380''.
}
\label{fig:hasta}
\end{center}
\end{figure}

%---------------------------------------
\subsection{Chromospheric Evolution}
\label{sec:obs_chromos} 

%{\S\bf~Global description}\\
An M4.7 flare was observed starting at 12:23 UT in GOES soft X-ray ligth-curve (SOL20120509T12:23:0) with maximum at $\approx$~12:32 UT. \fig{hasta} shows H$\alpha$ images from HASTA at six different times during the flare evolution. A movie starting at 11:55 UT and ending at 13:15 UT is included as electronic supplementary material (\href{run:./HASTA.mp4}{HASTA.mp4}). There is a change in the cadence of the data shown in the movie as the telescope changes to flare mode around the start time of the flare ($\approx$~12:23 UT). The data acquisition goes back to the normal cadence (an image every 2 to 4 minutes) at around 12:37 UT. 

%{\S\bf~H$\alpha$ evolution and difference images description}\\}\\
\fig{hasta}a shows the AR at approximately the time at which the flare started. Magnetic field contours are overlaid as a guide to the photospheric counterpart of the observed H$\alpha$ brightenings. To improve the visibility of the ``surge" in the HASTA data, we apply a cross-correlation technique to minimize the atmospheric jittering of the images. In order to increase the contrast of the moving material and any brightness variation in the data, we construct a base difference image sequence by subtracting from each image of the set the pre-event image obtained at 12:23:51 UT. In what follows we refer to the images processed in this way as base difference images. Panels b, e and f of \fig{hasta} are base difference images. We also include as supplementary material a movie containing the obtained sequence of base difference images (\href{run:./HASTA-diff.mp4}{HASTA-diff.mp4}).

\begin{figure}[]
\begin{center}
\includegraphics[angle=-90,width=\textwidth]{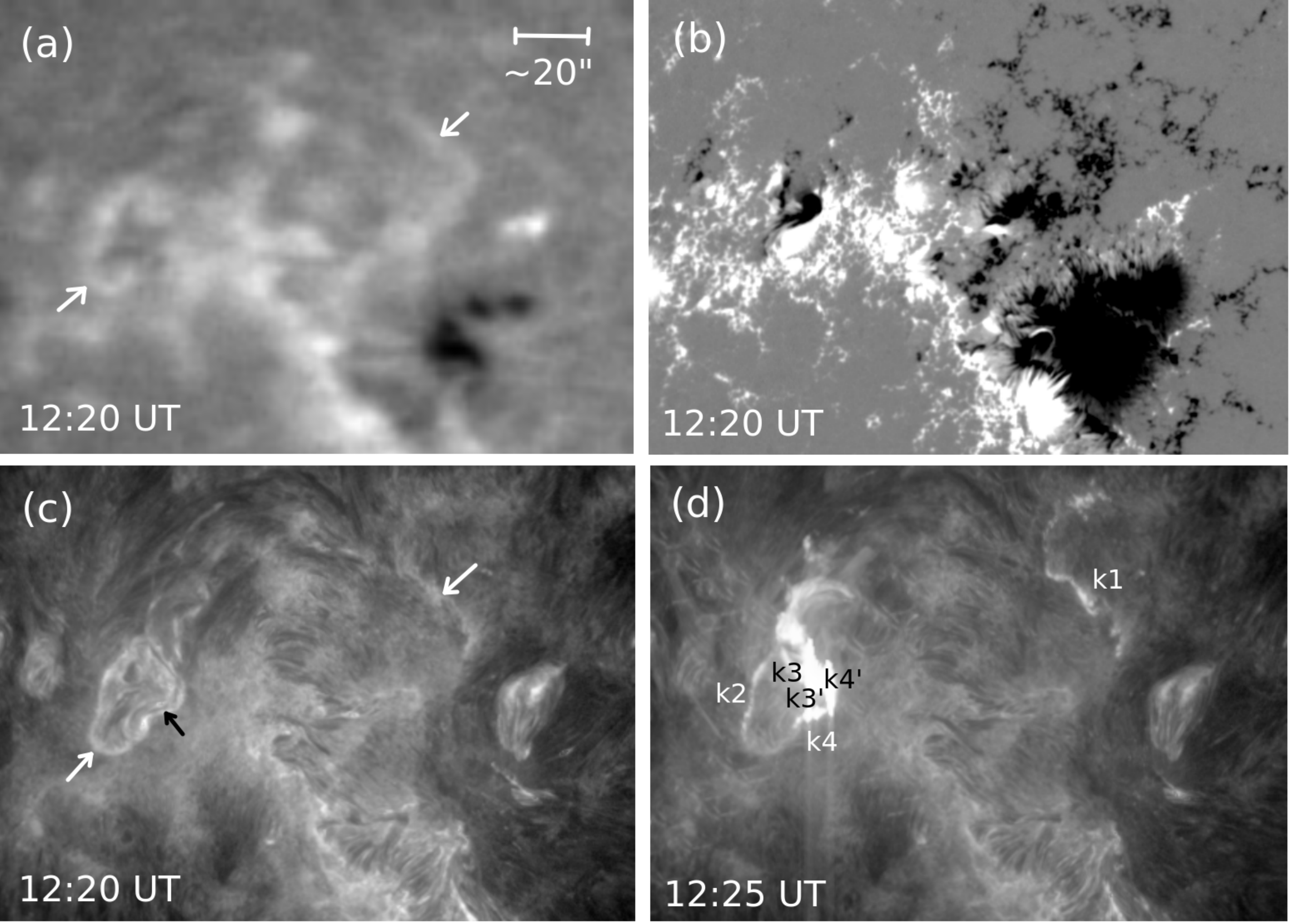}
\caption{(a) Enlarged view of the field covered by the gray box in \fig{hasta}a at a time %(12:20 UT) 
previous to the event. The white arrow on the left indicates the circular brightening partially surrounding bipoles B1 and B2 (see \sect{obs_B}). The white arrow on the right indicates a brightening on the main negative polarity of the AR. The white segment on top right of the panel is added to indicate the approximate scale size of the images. (b) Same field of view for HMI data for reference. (c) Same field of view for AIA 304 data (see also \fig{aia304}). The white arrows indicate the same corresponding features as in panel a, and the black arrow indicates the location of the minifilament. (d) AIA 304 image close to the event start time. Kernels k1 to k4, k3' and k4' are indicated. See also the accompanying movie included as supplementary material (\href{run:./surge-closeup.mp4}{surge-closeup.mp4}).}
\label{fig:closeup}
\end{center}
\end{figure}

\begin{figure}[]
\begin{center}
\includegraphics[width=0.9\textwidth]{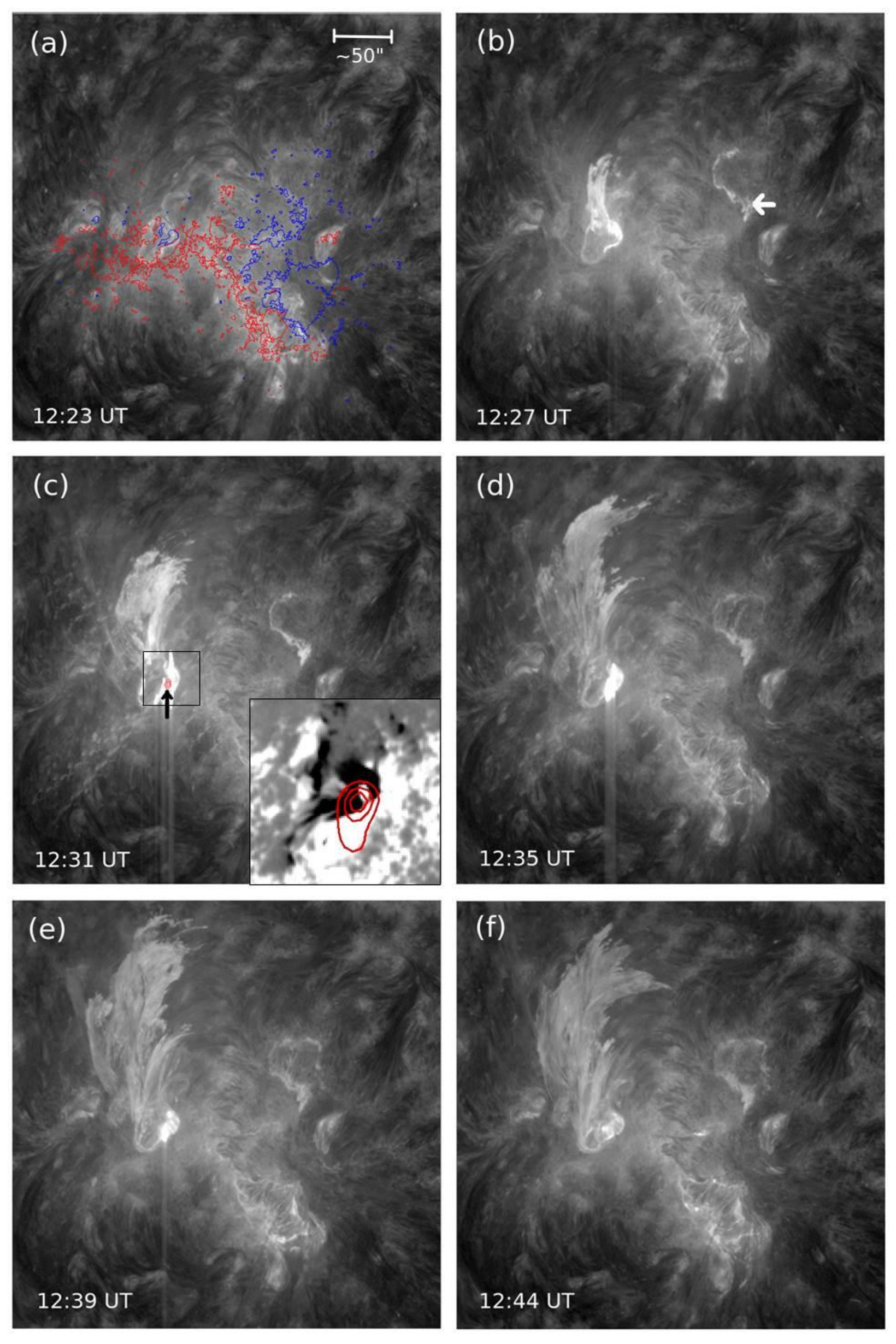}
\caption{SDO/AIA 304 images for the same times as those in \fig{hasta}. Times are indicated in the the bottom left of the panels. Panel a includes $\pm$ 300 G HMI contours as reference. Red (blue) corresponds to positive (negative) magnetic field. A white segment has been added on the top right of the panel a to indicate the approximate scale size. Panel c shows the location of RHESSI 40--80 keV contours (in red) at 40, 60, and 80\% of the maximum, integrated between 12:31:56 and 12:32:16 UT (around the time of the peak emission of the flare in soft X-rays). A small black arrow is added to facilitate viewing the location of the RHESSI contours. The inset in the lower right corner of this panel is a zoom of the rectangle that surrounds the brightest region in this AIA image. RHESSI contours in the inset are overlaid on the corresponding HMI magnetogram. A movie showing the AIA 304 evolution (\href{run:./AIA304.mp4}{AIA304.mp4}) is attached as electronic supplementary material. The square panels have a side length of approximately 380''. The inset in panel c covers approximately 50''$\times$50''. 
}
\label{fig:aia304}
\end{center}
\end{figure}

%{\S\bf~H$\alpha$ flare}\\
In \fig{hasta}b, we show a base difference image at $\approx$ 12:27 UT. In this panel, we identify an elongated brightening to the west overlapping part of the negative B1 polarity and indicating the presence of flare kernels. A less intense brightening is seen to the north-east (see the thick white arrow). This brightening elongates to the north and, when comparing this image and its corresponding movie with the AIA 304 observations of \fig{aia304}b and its corresponding movie, we conclude that it corresponds to heated ``surge" material seen as it starts moving upwards. As the flare developed (\fig{hasta}c) another elongated kernel became visible to the east of the first one, labeled k2, and a distant and extended one became prominent on the opposite side of the AR, labeled k1. These kernels are also present in AIA 304 images taken at the same times (\fig{aia304}). Comparing \figs{hasta}{aia304}, and according to the analysis of \sect{Processes}, we conclude that the brightening labeled as ``Flare kernels'' in \fig{hasta}b correspond to the overlapped emission of kernels k3', k4' and k3, while k4 is the smaller kernel at the south-east (see labels in \fig{hasta}c). A second series of brightenings (indicated with a white arrow in panel f) are observed in the south-west portion of the AR at later times (\fig{hasta}e--f) that are apparently not related to the studied flare.

\begin{figure}[]
\begin{center}
\includegraphics[width=0.9\textwidth]{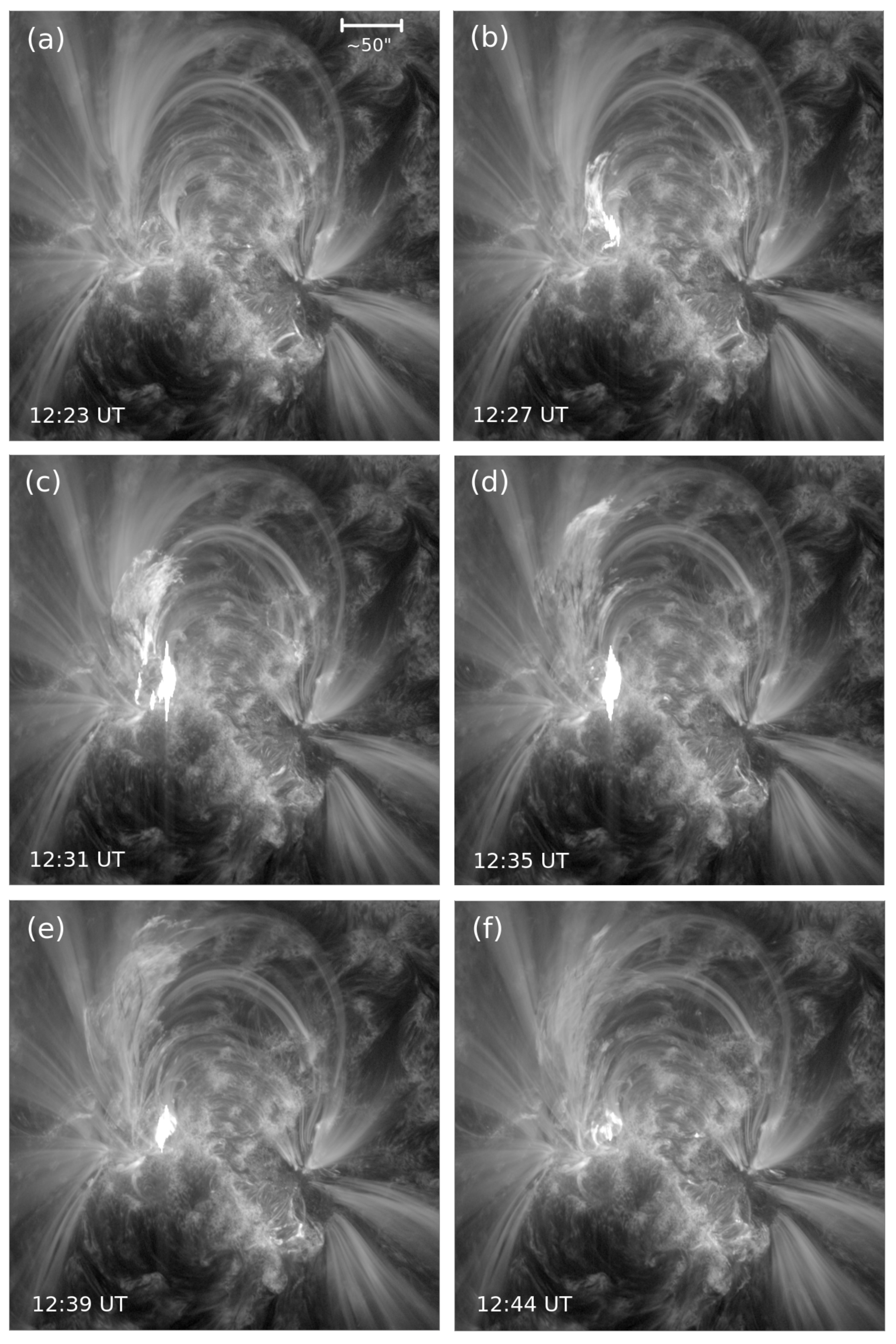}
\caption{SDO/AIA 171 images for the same times and fields of view as in \figs{hasta}{aia304}. The corresponding times are indicated in the bottom left of the panels. A white segment has been added on the top right of the panel a to indicate the approximate scale size. These images show the evolution of the EUV ``surge" and also part of the global structure of the AR as revealed by long EUV loops. A movie showing the AIA 171 evolution (\href{run:./AIA171.mp4}{AIA171.mp4}) is attached as electronic supplementary material.}
\label{fig:aia171}
\end{center}
\end{figure}

%{\S\bf~H$\alpha$ surge}\\
Although barely discernible, we identify in \fig{hasta}e--f the ``surge" (indicated with white arrows) as a dark feature in absorption at the north-east of the AR. The ``surge" can also be appreciated in absorption in the H$\alpha$ movies that accompany this article (\href{run:./HASTA.mp4}{HASTA.mp4} and \href{run:./HASTA-diff.mp4}{HASTA-diff.mp4}), where it is seen in motion. 

For a better identification of the H$\alpha$ features and their comparison with AIA observations, in \fig{closeup}a we show a zoomed image of the area indicated with a gray box in \fig{hasta}a at a time just before the event (12:20 UT). Notice the circular-shape brightening, pointed with a white arrow to the left, that surrounds the negative polarity of B1, part of its positive polarity, and all of B2, as described in \sect{obs_B}. This circular brightening is also observable in AIA 304 (also indicated with a white arrow in \fig{closeup}c) since around 08 May. The presence of this circular brightening strongly suggests that energy release at a very low rate was occurring accompanying the rotation of the bipoles probably because of their interaction with the overlaying magnetic field.  The identification of these kernels is confirmed by a close inspection of AIA 304 image shown in \fig{closeup}d, where they are labeled individually. k3' and k4' are the kernels associated to the footpoints of loops that result from the internal reconnection process thoroughly described in Section 3.3. Later on, the kernels extend and after the flare peak (12:32 UT) their intensity decreases.  

%---------------------------------------
\subsection{EUV Evolution from the Earth Point of View}
\label{sec:obs_euv}

%\quad{\S\bf~Figures description}\\
In AIA 304 images we also observe bright areas at the location of the H$\alpha$ plage regions, as expected. For a better comparison, in \fig{closeup}c we show the same area (indicated with a gray box in \fig{hasta}a) for AIA 304 before the event (12:20 UT). The white arrows indicate, respectively, the circular-shape brightening that surrounds the negative and part of the positive polarities of B1 and all of B2 (on the left) and the elongated distant kernel, k1 (on the right).

%\quad{\S\bf~minifilament}\\
As described in \sect{intro}, recent observations relate jets to the eruption of minifilaments which, as the word indicates, are smaller versions of normal large-scale filaments. High resolution observations as those presently obtained with imagers like SDO/AIA, allow identifying structures that were unobservable with the previous generation of solar instruments. \fig{closeup}c shows the presence of an elongated dark structure along B1 PIL compatible with the presence of a minifilament (see the black arrow in this panel). We also include as supplementary material a movie of this region (\href{run:./surge-closeup.mp4}{surge-closeup.mp4}) with a high temporal resolution (5 images {\it per} minute) showing the destabilization and eruption of the minifilament. In \sect{Processes} we analyze and discuss the role of this structure in the studied events.

%\quad{\S\bf~EUV surge}\\
\figs{aia304}{aia171} show AIA 304 and 171 images at approximately the same times as those of HASTA in \fig{hasta}. An HMI magnetogram contour is overlaid on the image in \fig{aia304}a as a reference. In \fig{aia304} panels b--e, as for the H$\alpha$ images, the main flare kernels are seen at the site of the main rotating bipole. The most striking feature seen in \figs{aia304}{aia171} is the dense EUV ``surge" moving towards the north-west part of the main negative AR polarity (numbered 1). This evolution starts right after the first appearance of the flare kernels. This is better seen in the corresponding movies that accompany the article as supplementary material (\href{run:./AIA171.mp4}{AIA171.mp4}, \href{run:./AIA304.mp4}{AIA304.mp4}, and \href{run:./surge-closeup.mp4}{surge-closeup.mp4}). The ejected material is seen to ascend fastly along the magnetic structure overlying the main bipole. This is clearer in the AIA 171 images in \fig{aia171}, where the AR loop structure before and during the event is well observed. 

%\quad{\S\bf~Velocity estimation from the 304 images}\\
We can make a rough estimation of the mean velocity of the EUV ``surge" front on the plane of the sky. Since the ``surge" motion is dominantly northward, we choose to follow the northward motion of the leading point of the front as a proxy for the ``surge" velocity projected on the plane of the sky. We obtain a value of $\approx$ 90 km s$^{-1}$, which agrees with the typical values measured in other ``surge" events (see references in \sect{intro}). 

%{\S\bf~Surge dynamics}\\
As observed in the AIA 304 and 171 movies, as soon as the EUV ``surge" reaches its maximum height part of the material drains towards the western footpoints of the loops along which it moves and part drains back to their eastern footpoints. The western loop footpoints, containing the draining plasma, end at the flare kernel k1, observed both in H$\alpha$ and AIA 304 images (\fig{hasta}c and \fig{aia304}b). The intensity of this region increases simultaneously with the flare. 

%{\S\bf~Surge apparent rotation}\\
As previously reported \citep[\eg,][]{Canfield96,Chae99} and also observed in the EUV movies that accompany this article (\href{run:./AIA171.mp4}{AIA171.mp4}, \href{run:./AIA304.mp4}{AIA304.mp4}), the upward motion of the ``surge" material has a torsional component. We associate the helicity accumulation to the precursory rotation of the bipolar structures described in \sect{obs_B} and its release in large-scale AR loops as due to the magnetic reconnection process associated to the flare. Although a rotational component is evident in the evolution of the ``surge" observed in AIA 171 and 304 data, the complex structure of the ejected material as projected on the plane of the sky makes it very difficult to obtain a definite answer regarding a sense of rotation and even more a quantitative estimation of this rotation.

%{\S\bf~Temporal evolution: H$_{\alpha}$ versus EUV}\\
Comparing the evolution of the ``surge" in H$\alpha$ and AIA 304 (\figs{hasta}{aia304}), the ``surge" seems to start with the ascent of material at the 304~\AA\ temperature range. We first see a bright elongation in HASTA images towards the north at $\approx$ 12:27 UT; the ``surge" is more prominent in AIA 304 at the same time, while it is only later that the material is observed barely expanding in absorption in H$\alpha$ when we compared the images in AIA 304 at the same times (\fig{hasta}e--f to \fig{aia304}e--f). Assuming that reconnection between the erupting minifilament and the overlying field starts at low coronal heights, we expect to see the just described evolution.  First, the plasma is heated up, rises, is seen bright in H$\alpha$ and dominantly in 304~\AA . Then, part of the ejected material cools down and becomes visible in absorption in H$\alpha$.  
 
%{\S\bf~Other EUV surge in the AR}\\
The presence of the circular-shape brightening around the bipoles and their sustained rotation, from before the event analyzed in this article and until more than a day after (10 May at around 22:00 UT for B1), motivated us to look for similar patterns of activity in this range of time. Using Helioviewer (\href{https://helioviewer.org/}{https://helioviewer.org/}), we are able to identify four extra EUV ``surges" in AIA 304, associated to M-class flares, one on 8 May at $\approx$13:05 UT (flare M1.4), another one on 9 May at $\approx$21:05 UT (flare M4.1), and two on 10 May at $\approx$04:15 UT \citep[flare M5.7, see, \eg,][]{Yang18} and at $\approx$20:25 UT (flare M1.7). All these flares and ``surges" started %by the region 
where the two rotating bipoles were located and the plasma evolved along large-scale loops with footpoints on the negative main AR polarity (numbered as 1) to the west, as is the case of the event studied in this article. This implies a recurrence of energy storage, probably minifilament rebuilding as proposed by \citet{Chandra17} \citep[see also the simulations of][]{Wyper17,Wyper18}, and energy release processes with a time scale between 7 and 23 hours and a mean of about 14 hours.

\begin{figure}[]
\begin{center}
\includegraphics[width=0.9\textwidth]{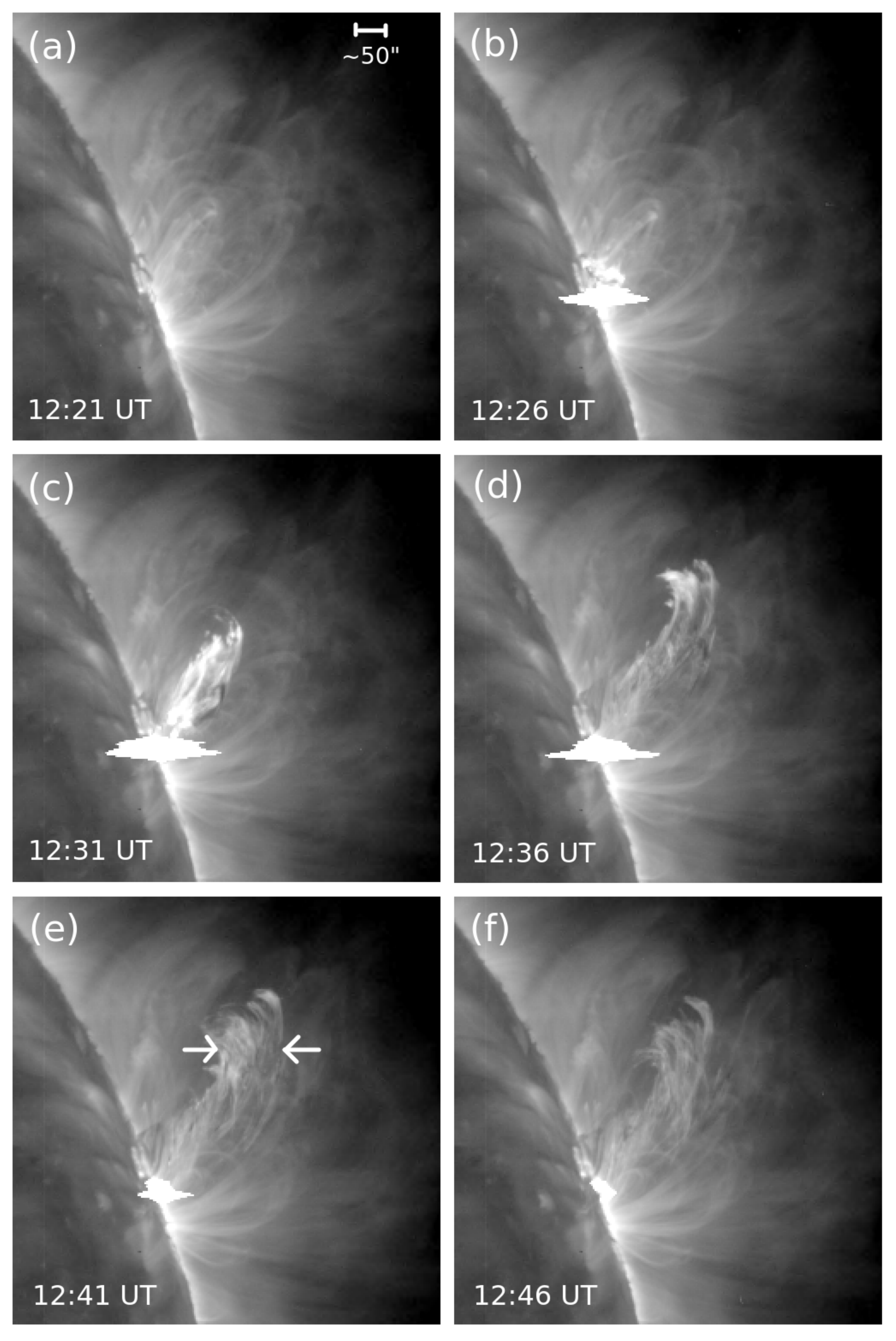}
\caption{STEREO-B/SECCHI images of AR 11476 in the 195 \AA~channel. At the date of observation the spacecraft was approximately located on the ecliptic at $118 \degree$ east of Earth. From this position the AR is observed on the solar limb. The saturated pixels correspond to the flare. The EUV ``surge'' evolution is clearly identifiable in the images. The observation times are provided in the bottom left of the panels and are the closest possible to those of \figss{hasta}{aia171}. A white segment has been added on the top right of the panel a to indicate the figure scale-size. A movie (\href{run:./STEREOB-195.mp4}{STEREOB-195.mp4}) showing the evolution of the EUV ``surge" in SECCHI images is attached as electronic supplementary material.}
\label{fig:stereo}
\end{center}
\end{figure}

%---------------------------------------
\subsection{EUV Evolution from STEREO-B Point of View}
\label{sec:obs_euv_STEREO}

%{\S\bf~STEREO-B data}\\
Complementing AIA data, we use SECCHI observations from STEREO-B, which at the time of the event was located towards the east solar limb as observed from Earth. From STEREO-B point of view, AR 11476 appeared on its northern right solar limb, as shown in \fig{stereo}. The panels of this figure correspond to different times in SECCHI 195 \AA~channel closest to the times of \figs{aia304}{aia171}. We use this wavelength because it has the highest available temporal resolution (5 minutes). We also produce a SECCHI 195~\AA~movie that covers the evolution from 12:00 UT to 14:55 UT (see \href{run:./STEREOB-195.mp4}{STEREOB-195.mp4} included as supplementary material). This movie and \fig{stereo} illustrate the evolution of the flare and EUV ``surge'' as observed at the limb by STEREO-B, above a series of saturated pixels because of the flare. 

%{\S\bf~Velocity observed from STEREO-B}\\
The vantage point of view of STEREO-B provides information, on the EUV ``surge'' structure, which is inaccessible from Earth's line of sight. For instance, an estimation of the ``surge" velocity using STEREO-B data gives $\approx$~120 km s$^{-1}$ in the plane of sky, which is consistent with typical velocity values cited in \sect{intro}. For this computation we follow the central point of the leading edge of the ``surge" seen against the dark and less dense background. The northward component of this motion is $\approx$~98 km s$^{-1}$. Considering that both spacecraft are located at Earth's ecliptic plane, this value is consistent with the 90 km s$^{-1}$ computed for the northward motion of the ``surge" as observed on the plane of the sky by SDO. The small difference is likely due to the geometrical (different plasma depth columns) and temporal differences (different instrumental cadences), errors involved, and the fact that observations correspond to different wavelength bands (304~\AA~ and 195~\AA) with different temperature responses. 

%---------------------------------------
\subsection{Estimation of Temperature and Density}
\label{sec:obs_euv_T}

%{\S\bf~Method}\\
The observed width of the EUV ``surge'' from a direction which is almost perpendicular ($\approx$118$\degree$) to SDO line of sight lets us also estimate the depth of the emitting material observed by AIA. We can use this information to make a rough calculation of the temperature and density of the ``surge" plasma by applying the so-called filter ratio technique to AIA 171 and 304 observed intensities. The procedure consists in assuming a quasi-uniform temperature for the ``surge" plasma. In this case, the intensity observed in a given AIA channel can be considered as the emission measure [$EM$] of the plasma multiplied by the instrument response function in that channel, $I_{ch1} = EM~S_{ch1}(T)$ \citep[see, \eg,][]{LopezFuentes07}. In this way, the ratio of the intensities in two different channels is a function only of the temperature, $R(T) = I_{ch1}/I_{ch2}$. Inverting this function we obtain $T$. Once the temperature is known, the density is computed from $n_{\rm e} = [EM/(f d)]^{1/2}$, where $f$ is the filling factor and $d$ is the depth of the observed plasma column. 

%{\S\bf~Results without subtracting the background}\\
We apply this procedure to the intensities averaged on the same corresponding ``surge" areas observed in AIA 171 and 304. We do this computation at the time when the EUV ``surge'' has its maximum apparent volume (at $\approx$ 12:39 UT). 
We determine a temperature of $2 \times 10^5$ K. Using the width of the EUV ``surge'' observed by SECCHI, estimated to be approximately 50 Mm (indicated with white arrows in \fig{stereo}e), as the depth of the plasma column [$d$] observed by AIA and setting $f = 1$, we obtain $n_{\rm e} = 1.8 \times 10^9 $ cm$^{-3}$. 

%{\S\bf~Results with background substracted}\\
We repeat the computation subtracting a mean of the background on both sides of the ``surge" for both AIA bands.  For AIA 304 the background is very small compared with the ``surge" intensity, but for AIA 171 the background can be as high as 80\% of the ``surge" intensity. Also, due to the presence of AR coronal loops the AIA 171 background tends to be particularly noisy in the vicinity of the surge. Therefore, different background choices provide a variety of temperature results. Trying different background combinations we obtain temperature values between 1.4 and 1.8$\times 10^5$ K and corresponding densities from 1 to 1.5$\times 10^9$ cm$^{-3}$. These values continue to be within the ranges measured in previously studied surge events (see references in \sect{intro}). A larger effect on the derived density comes from the assumed filling factor, since if $f \approx 0.1$ it implies a factor of $\approx 3$ on the computed density. Then, the obtained densities are lower limits if $f < 1$. 

%{\S\bf~Results with hotter filters}\\
Interestingly, although progressively less visible for hotter AIA bands, the ``surge" is still observable in AIA 193 and 211. Applying the filter ratio technique, using the same ``surge" and background selection as above, we obtain for the 193/171 combination $T = 1.5 \times 10^6$ K and for the 193/211 combination $T = 1.8 \times 10^6$. In both cases, $n_{\rm e} \approx 1 \times 10^8$ cm$^{-3}$, more than one order of magnitude smaller than for the 304/171 combination. This confirms that the bulk of the mass of the ``surge" emits at a temperature around the maximum AIA 304 response, while surprisingly the plasma pressure is comparable when using all ratio pairs.
  
\begin{figure}[]
\begin{center}
\includegraphics[width=0.9\textwidth]{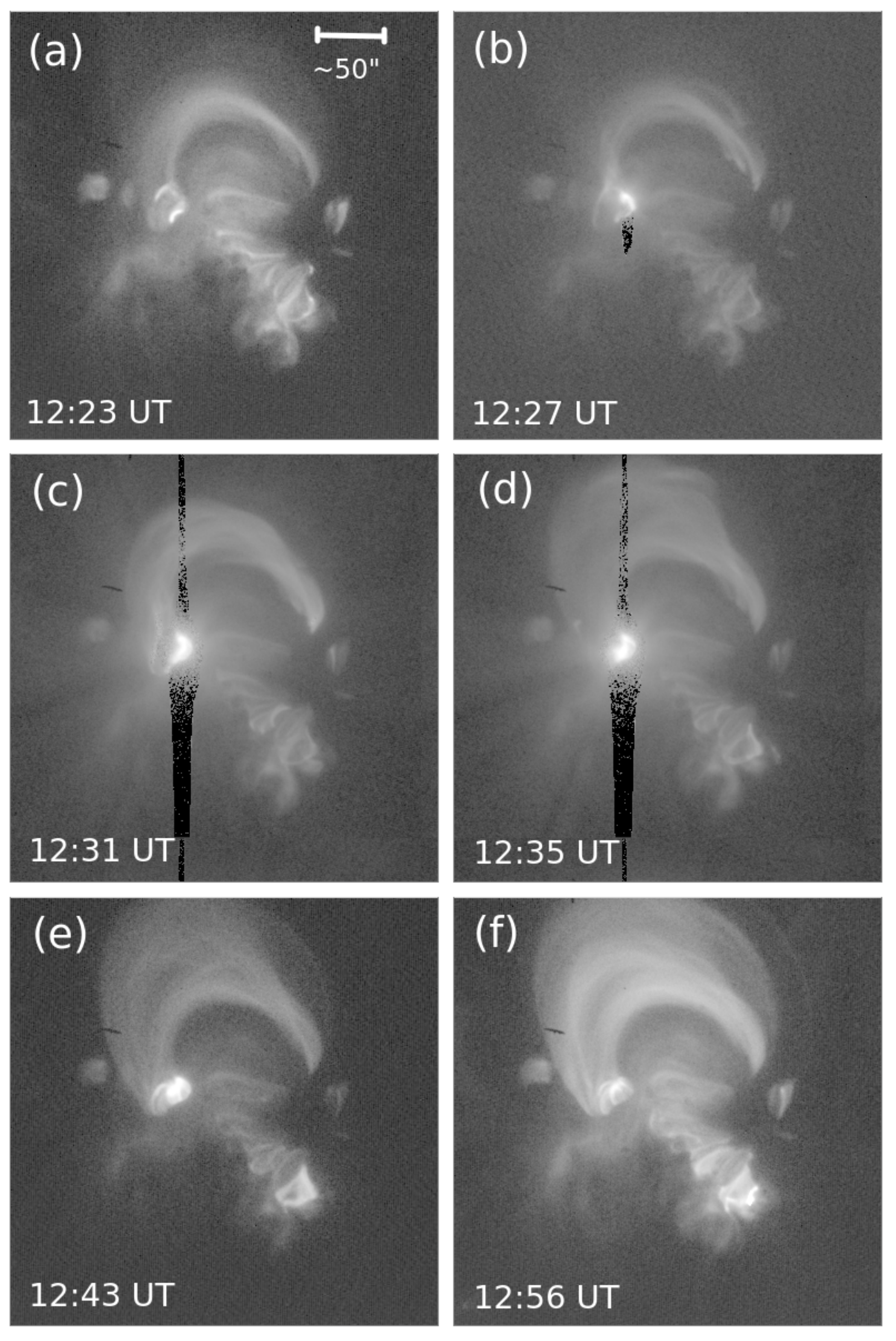}
\caption{{\it Hinode}/XRT images of AR 11476 for different times along the analyzed event evolution, but as close as possible to those of \figss{hasta}{aia171} and with the same field of view. The times are shown in the bottom left of the panels. A white segment has been added on the top right of the panel a to indicate the figure scale-size.} Images in panels b, c, and d are composites of images with short and long exposure times. The series of black pixels seen in these images are due to the effect of the composition. The bright structures associated with the flare are clearly seen in panels a to d. Their locations agree with those of the magnetic bipoles described in \sect{obs} and shown in \fig{tilt}.  A movie (\href{run:./XRT.mp4}{XRT.mp4}) showing the evolution of the large-scale loops in XRT images is attached as electronic supplementary material.
\label{fig:xrt}
\end{center}
\end{figure}

%---------------------------------------
\subsection{Hard X-Ray Evolution}
\label{sec:hxr_obs}

%{\S\bf~Hard X-ray observations}\\
RHESSI data in the 10\,--\,30 keV and 40\,--\,80 keV bands integrated during 20 s at the time of the flare peak ($\approx$12:32 UT) indicate the presence of hard X-ray emission at a very concentrated area close to the first H$\alpha$ elongated kernel (see \fig{aia304}c). A small black arrow is added to indicate the location of RHESSI contours in the AIA 304 image. The observed RHESSI emission is also close to the location of the minifilament (see \sect{obs_euv}) whose eruption was at the origin of the events described in this article. The inset in \fig{aia304}c shows (in red) an overlay of the RHESSI 40--80 keV contours at 40, 60, and 80\% of the maximum on a simultaneous HMI magnetogram. Notice that the estimated error in the location of RHESSI high energy source, coming from the processing procedure and coalignment between different instruments, is around 5\arcsec \citep{Lin02}. 

%{\S\bf~Effect of high energy particles on magnetogram}\\
The magnetogram in the inset shows a region of pixels displaying a magnetic field sign reversal with apparent positive (in white) polarity within the negative polarity of the main rotating bipole (compare with the magnetogram of \fig{hmi}, taken at 12:08 UT). This kind of anomaly has been observed in {\it Michelson Doppler Imaging} magnetograms at locations where beams of high energy particles precipitated at the maximum of some flares, see an explanation and the example shown by \citet{Qiu03} and another example in \citet{Mandrini06}.
Taking into account the uncertainty in the location of RHESSI sources, the field sign reversal region would agree with the highest intensity RHESSI contour.

%---------------------------------------
\subsection{Soft X-Ray Evolution}
\label{sec:sxr_obs}

%{\S\bf~XRT figure description}\\
\fig{xrt} shows the evolution of the flare observed with {\it Hinode}/XRT. Panels a--d of \fig{xrt} are for the same times as the corresponding ones in \figss{hasta}{aia171}. \fig{xrt}e and f show images at slightly later times (4 and 12 minutes later than the corresponding panels of \figss{hasta}{aia171}). An XRT movie showing the evolution from 12:17 UT to 13:52 UT is available as electronic supplementary material (\href{run:./XRT.mp4}{XRT.mp4}). The flare onset appears in the XRT images as an intensity increase in small elongated loop-like features at the location of the bipoles. Note in \fig{xrt}c--d the vertical set of dark pixels corresponding to the saturated area of the images that has been replaced with the corresponding pixels from images with shorter exposure times. Some of the pixels in these areas look black because the composite routine sets a value 0 to the pixels from the short exposure images that have intensities below the minimum of the long exposure images. Despite this, the bright loop-like features associated to the flare are clearly discernible in these images.

%{\S\bf~Loop evolution}\\
A set of longer loops connects the AR zone around the small bipoles to the western portion of the AR. These loops increase both in intensity and projected height as the extension and brightness of the flaring area increases (\fig{xrt}c--d).  The time at which this set of long loops reaches its maximum projected height coincides with the time at which the ``surge" observed in the EUV bands reaches its maximum height and apparent volume (around 12:39 UT).

%{\S\bf~Two sets of X-ray loops}\\
A few minutes after the soft X-ray intensity peak, during the gradual phase of the flare (\fig{xrt}e--f), two clear sets of loops are identifiable: short (low-lying) very bright loops at the location of the rotating bipoles and long loops that connect to the western part of the AR. The footpoints of these two sets of loops agree with the flare kernel locations in the H$\alpha$ images (\fig{hasta}) and in AIA 304 images (\fig{aia304}). The topological analysis of the AR structure (\sect{topology}) will help us understand the role of these observed structures.

%{\S\bf~Relation with the ``surge" structure}\\
Finally, it is also worth to mention that comparing the timing of the AIA 304 and XRT movies (\href{run:./AIA304.mp4}{AIA304.mp4}, \href{run:./XRT.mp4}{XRT.mp4}), the XRT set of large-scale loops have a much longer duration than the ``surge". They remain bright well after the ``surge" material has fallen back to the chromosphere.

\begin{figure}
\centering
\includegraphics[width=0.85\textwidth]{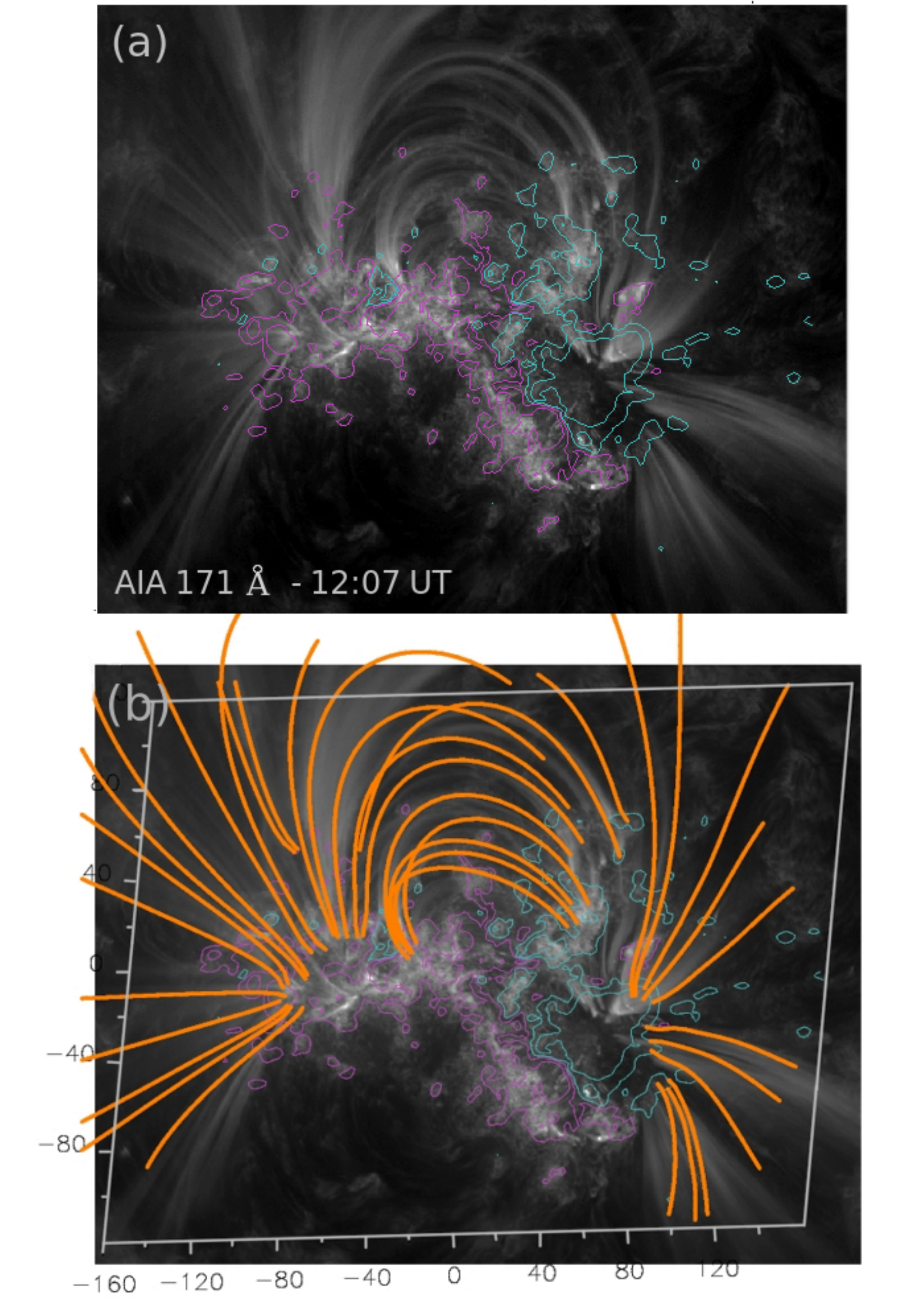}
\caption{(a) SDO/AIA 171 image at 12:07 UT overlaid with HMI isocontours with the same values as the ones in panel b, magenta (blue) for positive (negative) values of the field. (b) Coronal magnetic field model of AR 11476 overlaid on the 171~\AA\ image in panel a. Magnetic field lines (thick continuous orange lines) have been superimposed to be compared to the AR loops (see \sect{model} for the details on the model and $\alpha$ values). Isocontours of the field of $\pm$ 100, 500 G for positive (negative) values are shown in magenta (blue) color. The axis coordinates are in Mm in panel b; these are the units used in the computation of the model. The inclination of these axes indicates that the AR is slightly away from the disk center as seen from the perspective point of view of an observer at Earth. 
}
\label{fig:model_aia}
\end{figure}

%---------------------MODELING------------------------------------------

\section{Magnetic Field Topology of the Flare and EUV "Surge" Region}   
\label{sec:topology}

%---------------------------------------
\subsection{The Coronal Field Model}
\label{sec:model}

%\quad {\S\bf \quad The magnetic field model.}\\
To understand the role of the magnetic structures in AR 11476 during the flare and EUV ``surge", we first model its coronal field and later compute its magnetic topology. We extrapolate the HMI LOS magnetic field to the corona using the discrete fast Fourier transform method described by \citet{Alissandrakis81}, under the linear force-free field (LFFF) approach ($\curl \vec B = \alpha \vec B$, with $\alpha$ constant). 

%\quad {\S\bf \quad Practical settings of the model}\\
\fig{model_aia}a shows an AIA 171 image before the flare (12:07 UT) in which large scale magnetic loops are visible. 
\fig{model_aia}b displays the coronal model overlaid on the same AIA image. We use as the boundary condition for the model, the HMI magnetogram closest in time. We also apply a transformation of coordinates from the local frame (in which the computations are done) to the observed one so that our model can be compared to the flare brightenings and the EUV ``surge'' loops.
Although this model cannot take into account the distribution of currents at the photospheric level and the strong shear that the rotation of the bipoles could have created, but only the shear in the global magnetic configuration, it is fast and has proven to be efficient to compute the magnetic field topology to be compared with observed active events (see references in \sect{qsls}). Therefore, our magnetic field model and following topology computation represents the global AR magnetic structure (see \sect{qsls}).

%\quad {\S\bf \quad Set $\alpha$. Check its sign}\\
The value of $\alpha$, the free parameter of the model, is set to best match the observed loops \citep[as discussed by][]{Green02}. The best-matching value for the larger scale loops is $\alpha$ = -6.2 $\times$ 10$^{-4}$ Mm$^{-1}$, while for the shorter loops in the same image it is -3.1 $\times$ 10$^{-3}$ Mm$^{-1}$. This large-scale non-potentiality is related to the main AR polarities and not to the small rotating bipoles.  The AR appears on the eastern solar limb already emerged, therefore we cannot determine its age. However, its photospheric magnetic field, with strong sunspots that still have a distribution showing magnetic tongues (\fig{hmi}), indicates that it is a young AR close to its maximum flux stage.
The tongue-like pattern \citep[see][]{Poisson15} present for the main polarities of the AR is compatible with the sign of $\alpha$.

%---------------------------------------
\subsection{Quasi-Separatrix Layers}
\label{sec:qsls}

%\quad{\S\bf~Brief description of the QSLs theory and references}\\
\citet{Demoulin96} introduced the concept of quasi-separatrix layers (QSLs).
They are 3D thin volumes where the coronal field-line connectivity experiences a drastic change. QSLs are preferred sites for the formation of current layers and, therefore, locations where magnetic reconnection is expected to occur.  Several numerical experiments support
this idea \citep[see, \eg,][]{Milano99,Aulanier05,Buchner06,Pariat06,Wilmot-Smith09,Effenberger11,Savcheva12,Janvier13}. 

%\quad{\S\bf~Solar applications}\\
As concerns solar active phenomena, the computation of QSLs helps to interpret where flare kernels or other energy release manifestations should be observed \citep[\eg,][]{Demoulin97,Bagala00,Mandrini06,Mandrini14,Mandrini15,Cristiani07,Zhao14,Savcheva15,Janvier16,Polito17,Joshi17}. Furthermore, for moderately sheared or twisted magnetic structuring, the properties of QSLs in complex (multi-polar) configurations  depend strongly on the photospheric  distribution of the vertical field component and weakly on the details of the magnetic field model (\ie\ the spatial distribution of $\alpha$). In this sense, QSLs are a strong tool to understand where energy release proceeds during active phenomena and to learn about the properties of energy release sites \citep[see the reviews by ][]{Longcope05,Demoulin06,Mandrini10}. 

\begin{figure} %%%figure Qsls
\centering
\includegraphics[width=1.\textwidth, clip=]{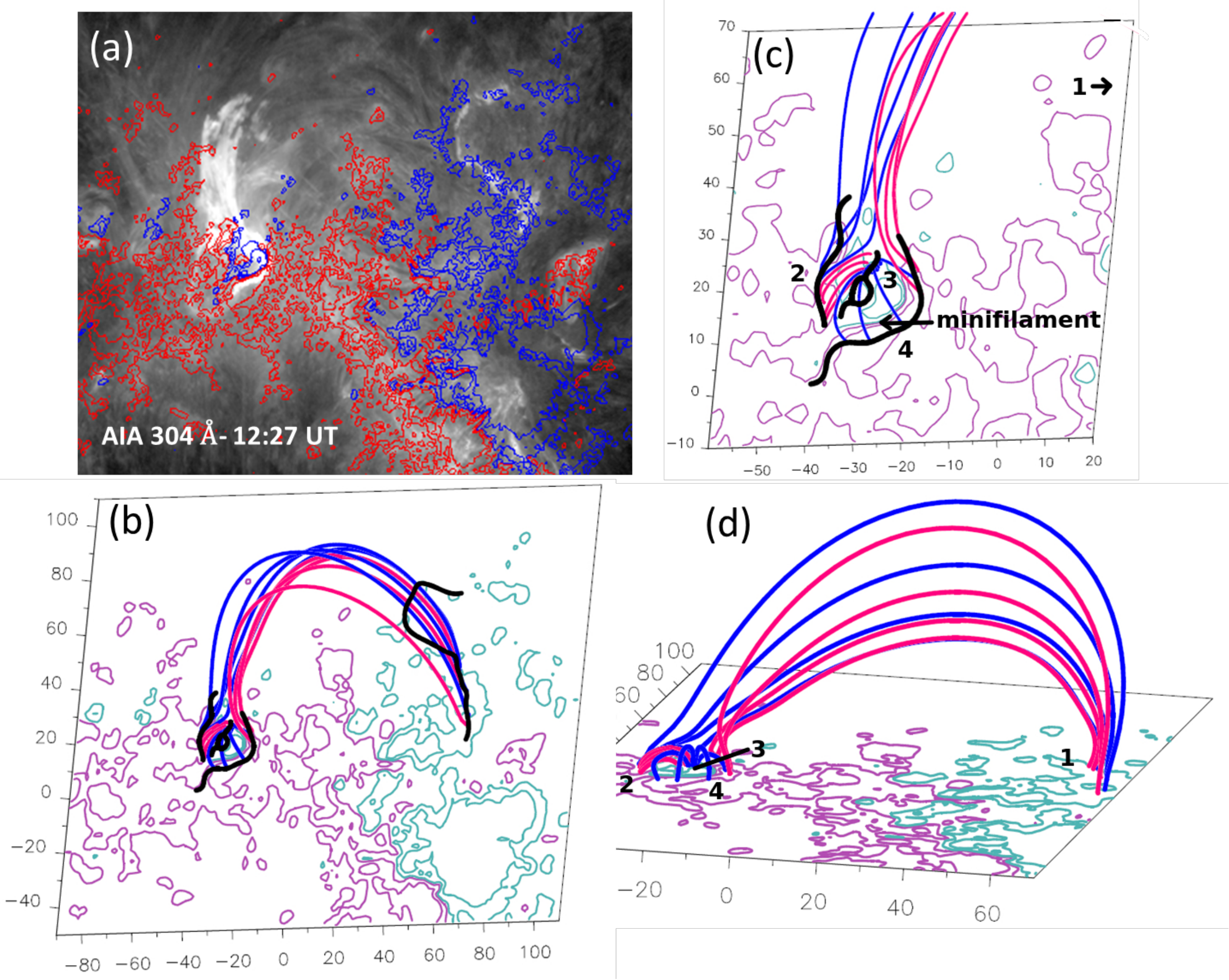}
\caption{
  (a) AIA 304 image at 12:27 UT overlaid by HMI isocontours. The flare kernels at the bipole locations and the elongated far kernel on the western negative AR polarity (numbered 1) are clearly visible. The EUV ``surge" is observed rising from eastern side of the bipoles.
  (b) Intersection of QSL traces (black thick continuous lines) with the photospheric plane for the same field of view as that shown in panel a. Two sets of field lines, representing the pre-reconnected (in blue) and reconnected field lines (in red), as inferred from the flare and ``surge" evolution, have been added (see also \fig{cartoon}). QSL traces match those of flare brightenings and the shape of the red field lines approximately follow the trajectory of the EUV ``surge".
  (c) A zoom of the QSL traces in the neighborhood of the bipoles whose rotation and cancellation forced the minifilament eruption and the initiation of the flare and EUV ``surge". An arrow points to the location of the minifilament along the PIL of bipole B1. Panels b and c are shown from the observer's point of view.
  (d) A different point of view selected to better display the 3D structure of the field lines. The axes in panels b, c, and d are in Mm and the isocontours of the field in all panels correspond to $\pm$ 100 and 500 G in continuous magenta (blue) style for the positive (negative) values.
The numbers 1--4 label the polarities mainly involved in the event and correspond to those in \figs{hmi}{cartoon} (see \sect{Processes}). The field of view of panel c does not include polarity 1, which is far to the west as indicated with the arrow.
}
\label{fig:qsl_conf}
\end{figure}

%\quad{\S\bf~Description of the procedure to compute the QSLs}\\
The original method to find the location of QSLs was described by \citet{Demoulin96}. QSLs were defined using the norm of the Jacobian matrix of the field-line mapping. In this method the value of the norm depends on the direction chosen to compute the mapping (from positive to negative photospheric polarities or the reverse). Then, the norm has different values at both footpoints of a field line on the photosphere. This ambiguity was solved by \citet{Titov02} who defined the squashing degree [Q], which is independent of the mapping direction. Q is the norm squared divided by the ratio of the vertical component of the photospheric field at both ends of a field line. In this way Q could be assigned to have a constant value along each field line. QSLs are computed by integrating extremely high numbers of field lines in a very precise way. To decrease the computation time we use an adaptive mesh to progressively refine the computation of field lines where Q has the highest values. We follow this iterative method until the QSL is locally well resolved or when the limit of the integration precision is reached \citep[see more details in ][]{Mandrini15}. 

%\quad{\S\bf~General description of figures}\\
As in other previous examples, we select large values of Q to show the extension of the QSL traces at the photospheric level. 
\fig{qsl_conf} displays the location of the intersection of QSLs with the photospheric plane. These photospheric traces are computed using the same HMI magnetogram shown in \fig{model_aia}, when the magnetic data do not show unrealistic values because of the injection of high-energy particles in the photosphere (see the inset of \fig{aia304}c). The results shown are computed with the highest $|\alpha |$ value, but similar results are obtained with the lowest $|\alpha |$ value (\sect{model}). QSL traces, shown as black continuous thick lines for simplicity and clarity, are overlaid on the photospheric magnetogram indicated as isocontours of the vertical component of the magnetic field. The value of Q for all QSL traces stays above 10$^{8}$, implying that these QSLs are very thin. 

%\quad{\S\bf~Comparison observations/model}\\
Panel a of \fig{qsl_conf} depicts an image of AIA 304 where the circular brightening at the location of the bipoles, the ``surge" at an early stage, and the far AIA 304 kernel, k1, on the main AR negative polarity, are observed.  Comparing with panel b, which is at the same scale, we see that the QSL photospheric traces agree well with the flare brightenings, while the EUV ``surge'' is cospatial with the set of large-scale red field lines (see also the ``surge" evolution in the movie \href{run:./surge-closeup.mp4}{surge-closeup.mp4}). \fig{qsl_conf}c shows a closer view of the shape of the QSLs around the rotating bipoles, as well as the QSL trace on the negative polarity of bipole B1 (numbered 3).  
Moreover, comparing panels a and b of \fig{qsl_conf}, the shape of the QSL trace far to the west matches the curved shape of the distant flare kernel, k1. Consequently, as in other analyzed examples, QSLs lie where flare kernels or other energy release manifestations are observed (see references at the beginning of this sub-section).

%\quad{\S\bf~Presence of null-points}\\
The distribution of the positive magnetic field around the negative polarity of B1, suggests the possibility of the presence of an associated  coronal magnetic null-point in the global AR magnetic field configuration \citep[see, \eg, the topology computations in][] {Mandrini14}. Furthermore, this is also the case for the almost circular shape of the QSLs and the observed brightening at the location of the rotating bipoles (see references in \sect{intro}). We searched for a magnetic null point in the model shown in \fig{model_aia} and did not find one that could be linked to the flare and ``surge". The reason why a null point is not present could be the fact that the negative B1 polarity is not completely surrounded by positive magnetic field (a negative extension is seen to the north of B1). However, no magnetic null-point and associated separatrices are needed in an observed configuration for energy release during an active event, as has been shown in many other examples (see references at the beginning of this section) or proposed in jet and/or surge models and simulations (see references in \sect{intro}). QSLs with a finite thickness are a generalization of the infinitely thin separatrices. If a null point is present its separatices surfaces will be found where QSLs are the thinnest and embedded within them \citep[see][]{Masson09,Masson17,Mandrini14}.    

%\quad{\S\bf~Small polarity}\\
Finally, we notice that the much smaller circular shape of the central QSL on the negative polarity of B1 (\fig{qsl_conf}c) appears because of a parasitic positive polarity that is present for a few hours at that location (this is a real parasitic polarity and not the artifact created by energetic particles, as shown in \fig{aia304}c). There is no flare brightening or activity related to this small polarity. Indeed, we claim that this parasitic polarity is not related to the flare and ``surge" studied in this article and we do not show the field line connectivity associated to it. 

%---------------------------------------
\subsection{Comparison of the Computed Topology to Observations}
\label{sec:Comparison-obs}
   
Panels b and d of \fig{qsl_conf} show a set of field lines computed from both sides of the circular QSL. Considering the observed evolution (\figss{hasta}{aia171}), we have drawn in blue field lines that represent loops before reconnection, while field lines in red correspond to loops after reconnection.
From the observation and model results, we conclude that the short blue field lines anchored at polarities labeled as 3 and 4 above the PIL of B1 reconnect with the long blue field lines anchored at 1 and 2.  A local zoom of the connectivities is shown in \fig{qsl_conf}c.  The two sets of red field lines are the result of this interaction. The shorter set is anchored at 2 and 3, where we observe flare kernels k2 and k3 in \figss{hasta}{aia171} at around 12:31 UT. The longer set follows the shape of the EUV ``surge" material moving to the west. This set of field lines starts at polarity 4, where kernel k4 is located, and ends at polarity 1, where the distant flare kernel k1 lies (compare panels a and b; see also panel d, which shows a more favorable view point, where we have not included the QSLs for clarity). 

Finally, when looking at the movie showing a close-up view of the minifilament eruption and ``surge" (\href{run:./surge-closeup.mp4}{surge-closeup.mp4}) one can see that the plasma first moves up along the longer set of red fields lines anchored at 1 and 4 and, by the end of the movie, it is seen to flow back towards the footpoints of the same field lines. Though, more difficult to see because the set of red field lines anchored at 2 and 3 is much shorter, a similar motion of the plasma along them is present. Finally, the field of view of the movie does not include the far kernel on the main AR polarity labeled as 1, but it is clearly observed in \figss{hasta}{aia304}, and the respective movies \href{run:./HASTA.mp4}{HASTA.mp4} and \href{run:./AIA304.mp4}{AIA304.mp4}.  

\begin{figure} %%%figure Cartoon
\centering
\includegraphics[width=\textwidth, clip=]{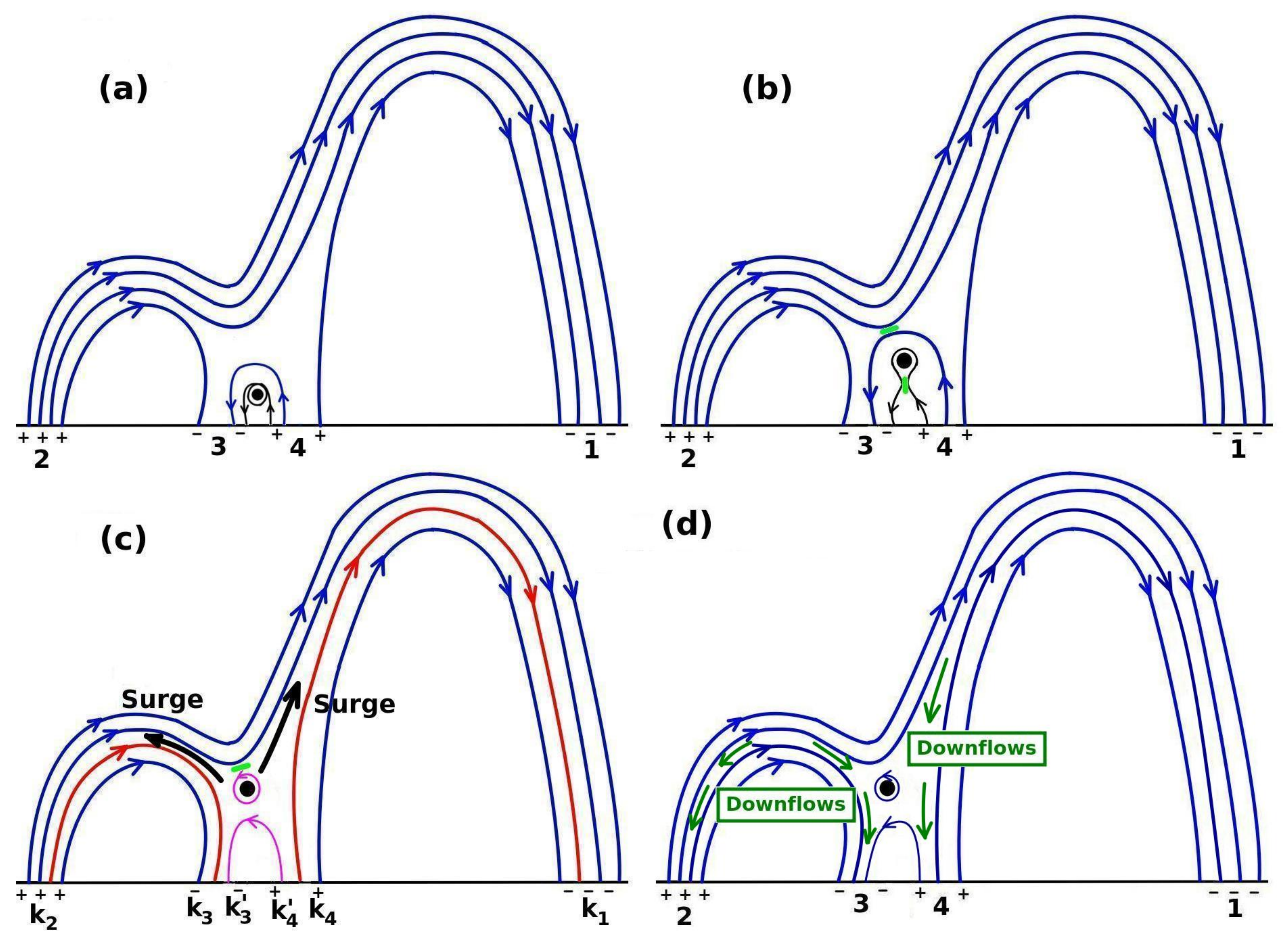}
\caption{2D Sketch illustrating the evolution of the flare and EUV ``surge''.  
  (a) Sets of blue magnetic field lines connecting polarities labeled 1--4.
The small black circle surrounded by a black field line represents the location of the minifilament. This panel corresponds to the magnetic configuration before the eruption. 
  (b) As the minifilament erupts two reconnection processes take place. Internal reconnection below the minifilament, indicated with a green short segment, and external reconnection, also indicated with a green segment above the minifilament. 
  (c) The two reconnection processes at work. Internal reconnection transforms the black field lines, located around the minifilament, in panel b in the pink field lines in this panel.
The result of this internal reconnection process is the appearance of flare kernels k3'and k4'. Simultaneously, external reconnection, between the blue lines connecting 3 and 4 and the longer blue lines connecting 1 and 2 in panel b, results in the red field lines connecting 2 with 3 and 1 with 4 indicated in this panel. The outcome of this second reconnection process is the injection of plasma forming the ``surge'' in the reconnected field lines. 
For a complete description of both reconnection processes, see \sect{Processes}.  
  (d) The downward flows are shown with dark green arrows. They occur later on in the previously reconnected field lines.     
}
\label{fig:cartoon}
\end{figure}

%---------------------------------------
\subsection{Physical Processes Involved}
\label{sec:Processes}

%\quad{\S\bf~Field line tracing in the figures}\\  
We now discuss a possible scenario to explain the observed events by analyzing in detail the connectivity of field lines traced starting integration at both sides of the QSLs. Since our connectivity analysis provides only a static view, to facilitate our discussion of the progression of the processes that occur from the beginning of the events analyzed in this article, we include the schematic drawing shown in \fig{cartoon}. This sketch is similar to others proposed to explain jets \citep[see, \eg,][]{Sterling15,Sterling16,Joshi18, Moore18} in open field configurations; however, in our example the background magnetic field is not open and unipolar but closed and bipolar.  

Panel a in \fig{cartoon} corresponds to the situation before the flare and ``surge". The numbers at the base of the field lines are labeling the magnetic polarities mainly involved in the events, as done in \fig{hmi}. 
The set of short blue field lines (anchored at 3--4) in \fig{cartoon}a overlays the minifilament that is located along the PIL of B1, while the set of long blue field lines (anchored at 1--2) constitute the background field, which is closed. 

Due, to a combination of flux cancellation along the B1 PIL and the bipole rotations, the minifilament destabilizes and starts rising (see \fig{cartoon}b). Magnetic reconnection sets in below the minifilament (indicated with a short thick green segment) as in a classical prominence eruption \citep[see \eg,][]{Aulanier10,Webb12}. The pink lines in \fig{cartoon}c correspond to the reconnected field lines resulting from this process, which has been called internal in several articles \citep[see \eg,][]{Sterling15, Moore18}.

Short blue field lines still restraining the minifilament start reconnecting with the large-scale blue lines (see \fig{cartoon}b and c). This second reconnection process, indicated with a short thick green segment above the minifilament, has been called external in the just mentioned references. The main difference with our particular case is that the configuration is closed.

%\quad{\S\bf~Failed eruption}\\  
In the case analyzed here, even if the magnetic tension above the minifilament could decrease as some of the overlying field lines are transformed by reconnection in the red field lines at both sides, anchored at 2--3 and 1--4 in \fig{cartoon}c \citep[see, \eg,][]{Antiochos99}, the plasma fails to reach farther out.  As a result, it is injected into these red reconnected field lines. The material is seen flowing along them forming a ``surge" (see \fig{cartoon}c). Finally, in \fig{cartoon}d we only indicate the direction of downflows along already reconnected lines as it occurs later on. To avoid an over busy panel, we omit any reconnection process in this panel. 

Summarizing, the first internal reconnection process would give the observed intense kernels at both sides of the PIL of B1, labeled k3' and k4' in \fig{closeup}d and \fig{cartoon}c, and very short loops joining them. This location roughly corresponds to the place of the RHESSI contours that indicate the deposition of accelerated particles (\fig{aia304}c). The second external reconnection process associated to kernels k1, k2, k3, and k4, observed in H$_{\alpha}$ and EUV (\figss{hasta}{aia304}), is what is captured by our global AR model and topology computation (\fig{qsl_conf}).

%---------------------------------------
\subsection{Comparison with Numerical Simulations}
\label{sec:Comparison-simu}

%\quad {\S\bf \quad Compare to simulations}\\
This observed case is close to the results of the simulations of \citet{Wyper17} and \citet{Wyper18} \citep[see also,][and references therein, though in these simulations no minifilament is formed by the imposed twisting motions]{Pariat16}. In fact, both global magnetic configurations (ours and that in the simulations) are comparable with an embedded polarity within the dominant polarity of opposite sign. The driver in both cases is the shearing of the arcade below the magnetic dome, in contrast with other cases where emergence is the dominant driver. Furthermore, in our observed case magnetic flux cancellation plays also a role.

There are still several differences between our observations and the simulations. A main difference is that the observed central polarity is not fully surrounded by the opposite sign polarity, which is plausibly why a magnetic null is not present in our extrapolation. From the QSL point of view, this does not imply a big difference since QSLs stay in place with a %variable 
thickness which is very thin in comparison with MHD scales. 

A second difference is the fully closed field configuration of our observations compared to the open field present in the simulations. Finally, the EUV ``surge" is much less collimated than the jet found in the simulations. Apart these differences, the underlying physical mechanism is the same: magnetic reconnection of a sheared arcade where the minifilament is embedded with a large scale field. 

%---------------------------------------
\subsection{Repetitive Events}
\label{sec:Repetitive}

%\quad {\S\bf \quad Repetition of the reconnection process}\\.
The above described processes started well before our analyzed example, as indicated by the fact that a ``surge" and flare with similar characteristics occurred on 8 May (\sect{obs_euv}). In fact, the storage then release of magnetic energy process is expected to continue for as long as the magnetic bipoles, their rotation, and field cancelation sustains. This evolution induces the reformation of a similar magnetic configuration including the minifilament.
 
Indeed three ``surges" and flares with similar characteristics occurred on 9 and 10 May after the one studied here.
Of course, the location of QSLs on the region of the rotating bipoles would slightly change following their evolution, but the main characteristics are expected to be kept with a circular QSL and a distant one, since this topology is robust for as long as the negative polarity exists within the positive surrounding. The same repetitive series of event was also found in some of the numerical simulations cited above \citep[][and references therein]{Pariat16}.

%-----------------CONCLUSIONS----------------------------------

\section{Conclusions}
\label{sec:conclusions}

%\quad{\S\bf~Summary of the events and computation of physical parameters}\\
In this article we analyze the solar flare and confined ejection that occurred in NOAA AR 11476 on 09 May 2012. For the analysis we use magnetic, H$\alpha$, EUV, soft and hard X-ray data from multiple space and ground-based instruments (see \sect{obs}). The way in which the plasma ejection develops is similar to the so-called blowout jets (see references in \sect{intro}). The main difference is that in the event studied here the material is ejected into closed magnetic field lines, while blowout jets usually occur (as regular jets do) along open magnetic structures. The velocity, angular span, and the fact that the ejection occurs in a closed magnetic structure reminds of classical surges, typically observed in H$\alpha$. However, it also bears differences with classical surges because a minifilament, whose plasma forms part of the ejection, is present in the configuration. Furthermore, as we show in \sect{obs}, the H$\alpha$ counterpart of the clearly observed ejection in AIA 304 and 171 is barely identifiable, that is why we call this confined ejection an EUV ``surge''. This is further confirmed by the thermal diagnostic analysis of \sect{obs_euv_T}, where we find that the ``surge" material has its maximum emission at an approximate temperature between 1.4 and 1.8$\times 10^5$ K, corresponding to the peak response of the AIA 304 channel.

Studying the evolution of the AR photospheric magnetic field during three days around the event, we identify two rotating bipoles in the central part of the main positive (following) polarity of the AR, that coincide in location with the flare and ``surge'' ejection. The magnetic flux of these bipoles decreases steeply during the period when the rotation is observed. A minifilament is seen lying along the PIL of the largest rotating bipole. We conclude that the sustained rotation of the bipoles is the main source of helicity and free magnetic energy accumulation that is later released during the events. 

We estimate a northward ``surge" velocity of 90 km s$^{-1}$ on the plane of the sky as seen from Earth, which is consistent with the velocity estimation of 120 km s$^{-1}$ as observed from STEREO-B point of view (see \sects{obs_euv}{obs_euv_STEREO}). These are typical velocity values observed in previously studied jets and surges (see \sect{intro}). We also identify a torsional component in the development of the ejection, which is very likely associated to the magnetic helicity injected during the rotation of the bipoles. 

%\quad{\S\bf~Role of the minifilament}\\
As in several recent observations of coronal jets, a minifilament is present along the PIL of the main rotating bipole. We propose that its rotation, combined with flux cancellation, destabilizes the minifilament and brings it to eruption. Magnetic reconnection below the minifilament, called internal reconnection, is mainly responsible for the brightest flare kernels and particle acceleration whose deposition sites are seen as elongated RHESSI contours. Simultaneously, external magnetic reconnection occurs between the loops within which the erupting minifilament is embedded and the closed AR loops forming the background field. This reconnection is at the origin of four flare kernels. Two are located on the main rotating bipole while the two others are located on the AR polarities.

%\quad{\S\bf~Model, topology more global}\\
Next, we combine the observations with coronal magnetic field extrapolations using HMI magnetograms as boundary condition (\sect{model}). Our global magnetic field model lets us explain the external reconnection process occurring as the minifilament and its field configuration are destabilized and erupted. We determine the location of the QSL traces at the photospheric level and relate them with the observed location of flare kernels (\sect{Processes}). In particular, the shapes and locations of QSL traces consist of an almost circular QSL surrounding most of the two rotating bipoles plus an elongated QSL on the main AR negative polarity to the west (see \fig{qsl_conf}). The circular QSL trace coincides with the flare kernels observed both in H$\alpha$ and EUV, while the elongated one corresponds to a far flare kernel (\sect{Processes}). 

%\quad{\S\bf~Faint energy release and buildup }\\
Furthermore, a continuous faint brightening at the location of the circular QSL is observable about a day before our analyzed event, showing that the above topology was present long before.  We argue that the sustained bipole rotations ($\approx$ 72 hs) 
allows for magnetic helicity and energy storage between, at least five flare and ``surge" events (one on 8 May and two on both 9 and 10 May) and also, most probably, contributes to the reformation of the minifilament. We find that the time scale of energy storage (with slow reconnection involved) is between 7 and 23 hours, with a mean of about 14 hours.

%\quad{\S\bf~Field line and reconnection}\\
Going deeper in the analysis, we select as a starting integration points both sides of the computed QSLs to compute sets of magnetic field lines with different connectivities and reconstruct the spatial evolution of the event, which allows us to determine the role of the different loop structures. The computed connectivities confirm the above interpretation, as summarized in \fig{qsl_conf}.  This allows us to derive a detailed scenario of the physics involved (\fig{cartoon}).

%\quad{\S\bf~What is remarkable: no emergence}\\
 What is remarkable in the events analyzed here is that, contrary to what has been proposed in numerical jet and surge simulations (see references in the \sect{intro}), the ``surge" and flare do not seem to have been driven by magnetic flux emergence. 
This has been also shown to be the case in other jet observations \citep[see \eg,][]{Adams14,Panesar16,Joshi18}.
The magnetic flux measurements indicate that the bipoles (mainly B1) are in decay, strongly suggesting that the main driver is the shearing of the arcade and flux cancellation within bipole B1.

%\quad{\S\bf~Compare to simulations}\\
This observed example is similar to the results of the simulations by \citet{Wyper17} and \citet{Wyper18} \citep[see also,][and references therein]{Pariat16}. The magnetic configurations, observed in our case and simulated in the previous articles, are comparable, since both have an embedded polarity within a dominant polarity of opposite sign. The driver in both cases is due to the shearing of an arcade, with flux cancellation also playing a role in the filament destabilization. A difference is that our configuration is formed by closed field lines compared to the open field in the simulations and, furthermore, the EUV ``surge" is less collimated than the jet found in the numerical model. 

%\quad{\S\bf~Closing}\\
Finally, the global AR magnetic field structure analyzed here lets us conclude that no magnetic null-point and associated separatrices are needed in an observed configuration for energy release, as has been shown in many other examples (see references in \sect{qsls}). QSLs with a finite thickness are a generalization of the infinitely thin separatrices. If a null point is present its separatices are located where QSLs are the thinnest and embedded within them \citep[see][]{Masson09,Masson17,Mandrini14}.

%%%%%%%%%%%%%%%%%%%%%%%%%%%%%%%%%%%%%%%%%%%%%%%%%%%%%%%%%%%%%%%%%%%%%%%%%%%
%% Appendix

%\appendix
%\section{}
%\label{sec:}

%%%%%%%%%%%%%%%%%%%%%%%%%%%%%%%%%%%%%%%%%%%%%%%%%%%%%%%%%%%%%%%%%%%%%%%%%%%
%% Acknowledgements
%
\begin{acks}

The authors are grateful for the insightful referee who stimulated us to improve several parts of the manuscript.
The authors would like to thank Dr. Carlos Francile from the Astronomical Observatory Felix Aguilar, University of San Juan, for his invaluable help in the processing of HASTA data. MLF, CHM and GC are members of the Carrera del Investigador Cient\'{\i}fico of the Consejo Nacional de Investigaciones Cient\'{\i}ficas y T\'ecnicas (CONICET) of Argentina. MP and FL are CONICET Fellows. MP, MLF, GC and CHM acknowledge financial support from the Argentinean grants PICT 2012-0973 (ANPCyT), UBACyT 20020130100321 and PIP 2012-01-403 (CONICET).

\end{acks}

\section*{Disclosure of Potential Conflicts of Interest}
The authors declare that they have no conflicts of interest.

%%% %%%%%%%%%%%%%%%%%%%%%%%%%%%%%%%%%%%%%%%%%%%%%%%%%%%%%%%%%%%
%% Bibliography
%
% Using BibTeX
%
% \bibliographystyle{spr-mp-sola}
% \bibliography{<bib file>}  

%
% Without BibTeX 
% \begin{thebibliography}{}
% \bibitem[\protect\citeauthoryear{Author}{Year}]{key}
%   <bibliographical entry>
%
% \bibitem[\protect\citeauthoryear{}{}]{}
%   

%  
% \end{thebibliography}

\bibliographystyle{spr-mp-sola} % format of ref. provided by the review (.bst)
\bibliography{paper_surge_astro-ph} % file containing the bibtex references (.bib)
      % look if the file containing the ``\bibitem'' exits
\IfFileExists{\jobname.bbl}{}
{\typeout{}
\typeout{****************************************************}
\typeout{****************************************************}
\typeout{** Please run "bibtex \jobname" to optain}
\typeout{** the bibliography and then re-run LaTeX}
\typeout{** twice to fix the references!}
\typeout{****************************************************}
\typeout{****************************************************}
\typeout{}
}

\end{article} 

\end{document}